\documentclass[preprint,amsmath,amssymb,aps,pre,nobibnotes,tightenlines]{revtex4-1}
\usepackage{amssymb,latexsym,amsmath,amsxtra,mathtools}
\usepackage{amsfonts}
\usepackage{xspace}
\usepackage[final]{graphics}
\usepackage{color,psfrag} 
\usepackage[labelformat=simple]{subfig}
\usepackage{enumitem}
\usepackage{siunitx}
\usepackage[colorlinks,bookmarks,urlcolor=blue,citecolor=blue,linkcolor=blue]{hyperref}

\usepackage{bbm}

\renewcommand{\vec}[1]{\boldsymbol{#1}}

\newcommand{\paren}[1]{\left(#1\right)}
\newcommand{\brac}[1]{\left[#1\right]}
\newcommand{\E}{\mathbb E}
\newcommand{\avg}[1]{\E[#1]}

\newcommand{\D}[2]{\frac{d#1}{d#2}}
\newcommand{\DD}[2]{\frac{d^{2}{#1}}{d{#2}^{2}}}

\newcommand{\lap}[1]{\Delta#1}

\newcommand{\abs}[1]{\left|#1\right|}

\DeclareMathOperator{\prob}{Pr}

\newcommand{\kb}{k_{\text{B}}}

\newcommand{\bi}{\vec{i}}
\newcommand{\bj}{\vec{j}}

\newcommand{\ve}{\vec{e}}

\newcommand{\vxi}{\vec{\xi}}

\newcommand{\bib}{\bi_{\textrm{b}}}

\def\R{\mathbb{R}}

\newcommand{\Z}{\mathbb{Z}}

\newcommand{\phimax}{\phi_{\textrm{max}}}
\newcommand{\Pss}{P^{\textrm{ss}}}

\newcommand{\Pin}{P^{\,\textrm{in}}}
\newcommand{\Pout}{P^{\,\textrm{out}}}
\newcommand{\CO}{C_{\vec{0}}}
\newcommand{\Teuc}{\tau_{\textrm{euc}}}
\newcommand{\comment}[1]{{#1}}

\begin{document}

\title{The Influence of Spatial Variation in Chromatin Density
  Determined by X-ray Tomograms on the Time to Find DNA Binding Sites}


\author{Samuel A. Isaacson}
\email{isaacson@math.bu.edu}
\affiliation{Department of Mathematics and Statistics, Boston University, Boston, MA}

\author{Carolyn A. Larabell}
\email{Carolyn.Larabell@ucsf.edu}
\affiliation{Department of Anatomy, University of California, San Francisco, CA}
\affiliation{Lawrence Berkeley National Laboratory, Berkeley, CA}

\author{Mark A. Le Gros}
\email{MALeGros@lbl.gov}
\affiliation{Department of Anatomy, University of California, San Francisco, CA}
\affiliation{Lawrence Berkeley National Laboratory, Berkeley, CA}

\author{David M. McQueen}
\email{mcqueen@cims.nyu.edu}
\affiliation{Courant Institute of Mathematical Sciences, New York University, New York, NY}

\author{Charles S. Peskin}
\email{peskin@cims.nyu.edu}
\affiliation{Courant Institute of Mathematical Sciences, New York University, New York, NY}



\begin{abstract}
  In this work we examine how volume exclusion caused by regions of
  high chromatin density might influence the time required for
  proteins to find specific DNA binding sites. The spatial variation
  of chromatin density within mouse olfactory sensory neurons is
  determined from soft X-ray tomography reconstructions of five
  nuclei.  We show that there is a division of the nuclear space into
  regions of low-density euchromatin and high-density heterochromatin.
  Volume exclusion experienced by a diffusing protein caused by this
  varying density of chromatin is modeled by a repulsive potential.
  The value of the potential at a given point in space is chosen to be
  proportional to the density of chromatin at that location. The
  constant of proportionality, called the volume exclusivity, provides
  a model parameter that determines the strength of volume exclusion.
  Numerical simulations demonstrate that the mean time for a protein
  to locate a binding site localized in euchromatin is minimized for a
  finite, non-zero volume exclusivity.  For binding sites in
  heterochromatin, the mean time is minimized when the volume
  exclusivity is zero (the protein experiences no volume exclusion).
  An analytical theory is developed to explain these results. The
  theory suggests that for binding sites in euchromatin there is an
  optimal level of volume exclusivity that balances a reduction in the
  volume searched in finding the binding site, with the height of
  effective potential barriers the protein must cross during the
  search process.
\end{abstract}


\maketitle

\section{Introduction}

What mechanisms drive the process by which regulatory proteins and
transcription factors search for specific DNA binding sites? It is
usually assumed that in the absence of any interactions, within an
``empty'' nucleus a protein will move by Brownian motion and bind upon
getting sufficiently close to a target site. This corresponds to the
well-known diffusion-limited reaction model of
Smoluchowski~\cite{SmoluchowskiDiffLimRx}. A number of theoretical
and experimental studies suggest proteins may experience facilitated
diffusion, allowing them to find specific binding sites faster
than the diffusion limit predicted by the Smoluchowski
theory~\cite{HalfordBindRatesAreDiffLim09,NormannoSearchRev2012}.
Whether facilitated diffusion occurs in any meaningful way \textit{in
  vivo}, allowing proteins to find binding sites significantly faster
than the diffusion limit, is very much an active area of investigation
and
debate~\cite{HalfordBindRatesAreDiffLim09,NormannoSearchRev2012,NudlerNatStruct2013,VekslerSpeedSelect2013}.
In the present paper, we consider the possible influence of volume
exclusion caused by spatial heterogeneity in chromatin density on the time
needed for a protein to find a target by diffusion.

\comment{There are many different interactions that have been proposed
  that could in principle decrease the mean time required for
  regulatory proteins and transcription factors to find specific DNA
  binding sites, relative to the diffusion limited reaction model.}
These include non-specific DNA binding interactions, electrostatic
interactions between proteins and binding sites, one-dimensional
diffusion of proteins along DNA, and jumping of proteins between
different regions of DNA
fibers~\cite{HalfordBindRatesAreDiffLim09,NormannoSearchRev2012}.  The
relative contribution of these (possible) interactions is still being
assessed by both experimental and theoretical studies. Perhaps the
most popular of these mechanisms is the possibility that proteins may
exploit non-specific DNA binding interactions to allow diffusion
\comment{or sliding} along DNA.  In the classic study of Berg et
al~\cite{BergSlidingI}, a model was developed in which proteins could
undergo a mixed search process involving periods of three-dimensional
diffusion, coupled to periods of one-dimensional diffusion along DNA
fibers when the proteins were non-specifically bound.  A large number
of theoretical studies have investigated how this mechanism might
influence the search process for specific DNA binding sites (for
example, see~\cite{BergSlidingI, ElfSingMolSlidingRoadBlocks09,
  HolcmanDNASlide08, MirnyDNASlide04, MirnySlideGlob09,
  HalfordBindRatesAreDiffLim09}). Recently, several single-molecule
imaging studies have demonstrated that sliding of \textit{lac}
repressor can occur \textit{in
  vivo}~\cite{ElfSingMolSliding07,ElfScience2012}. The theoretical
estimates in~\cite{ElfScience2012} suggest this mechanism may allow a
significant decrease in the mean time required for a protein to find a
specific binding site, \comment{in comparison to the diffusion limited
  reaction model}.  Most of the existing studies have focused on
prokaryotic cells, and it remains to be seen whether sliding along
chromatin in eukaryotic cells can noticeably reduce the time required
for regulatory proteins to locate specific binding sites \textit{in
  vivo}.  More complete references for both theoretical models and
previous experimental work can be found in the
reviews~\cite{HalfordBindRatesAreDiffLim09,NormannoSearchRev2012}.

The nucleus of a eukaryotic cell is a complex spatial environment,
containing chromatin fibers with spatially-varying compaction levels,
nuclear bodies, and fibrous filaments (such as the nuclear lamina).
Spatial inhomogeneity of the nuclear space provides another possible
mechanism that may influence the search process of proteins for
specific binding sites. In~\cite{EllenbergVolExEMBOJ09} the impact of
spatial variation in chromatin density on the movement of proteins
within the nucleus was investigated.  Using a combination of
single-particle tracking experiments, photo-activation experiments and
computational modeling, the authors concluded that chromatin dense
regions, such as heterochromatin, exhibited noticeable volume
exclusion compared with less dense regions, \comment{such as
  euchromatin}.  It was observed that similar fluorescence activation
curves were found in photo-activation experiments within
heterochromatin and euchromatin, \comment{when each of the two curves
  was normalized to the steady-state fluorescence level of its own
  region}.  The authors inferred from these experiments that
heterochromatin is not substantially more difficult for proteins to
enter than euchromatin, but that heterochromatin has a smaller amount
of free space in which proteins can accumulate.  In contrast, the
supplemental movies of~\cite{RajMRNADiffPNAS2005} demonstrate that
individual mRNAs that are able to move freely within nuclei appear
restricted to regions of low histone-GFP fluorescence. \comment{These
  mRNAs seem to have difficultly moving into regions of
  heterochromatin, as identified by regions of high histone-GFP
  fluorescence.}

In~\cite{IsaacsonPNAS2011} we developed a mathematical model to
investigate how the spatially varying density of chromatin within
eukaryotic cell nuclei might influence the time required for proteins
to find specific binding sites. Our model assumed that regions of
higher chromatin density were more difficult for proteins to move
into. Protein motion was approximated as diffusion within a
volume-excluding potential.  The model was constructed from the 3D
structured illumination microscopy fluorescence imaging data
of~\cite{GustafsonScience08}. In that work, mouse myoblast cell nuclei
were chemically fixed, and both antibody labeled nuclear pores and
DAPI stained DNA were imaged.  From these data we reconstructed a
nuclear membrane surface to determine the nuclear space.  The
normalized DAPI stain intensity within a given voxel of the imaging
data was assumed proportional to the density of chromatin within that
voxel. Based on this assumption we constructed a volume exclusion
potential, with the value of the potential within a given voxel chosen
to be proportional to the normalized DAPI stain intensity of that
voxel.  The constant of proportionality, which we called the
\emph{volume exclusivity}, was a model parameter that set the overall
strength of volume exclusion. By varying the volume exclusivity, we
studied how the time to find specific DNA binding sites varied when
there was no volume exclusion (\textit{i.e.} the volume exclusivity
was zero and the protein simply diffused), weak volume exclusion, and
strong volume exclusion from chromatin dense regions.  Numerical
simulations of the protein's search process suggested that for binding
sites localized in regions of low DAPI stain intensity, such as the
20th to 30th percentile of intensity values, the median time for
proteins to find a specific binding site was minimized for non-zero
values of the volume exclusivity. That is, as the volume exclusivity
was increased from zero the median binding time initially decreased to
a minimum, beyond which the median time increased to infinity. After
randomly shuffling the values of the DAPI stain intensity among the
voxels of the nucleus, we observed that this effect was lost and the
median binding time simply increased as a function of the volume
exclusivity. Based on these results, we concluded that the spatial
organization of chromatin played a role in the observed minimum of the
median binding time for non-zero volume exclusivity.  For binding
sites localized in regions of high DAPI stain intensity, such as the
70th to 80th percentile of the DAPI stain intensity distribution, the
median time to find the binding site increased monotonically as the
volume exclusivity was increased from zero.

In this work we expand upon the studies begun
in~\cite{IsaacsonPNAS2011}. To determine chromatin density fields we
now use 3D soft X-ray tomography (SXT) reconstructions of cell
nuclei~\cite{McDermott:2009ib,LarabellCell2012,LarabellCell2013}.  SXT
provides several advantages over fluorescence imaging in assessing the
spatial variation of chromatin density. Foremost, the measured linear
absorption coefficient (LAC) of a voxel within SXT reconstructions is
linearly related to the density of organic material within that voxel
by the Beer-Lambert Law~\cite{McDermott:2009ib}.  Our simulations
in~\cite{IsaacsonPNAS2011} made use of imaging data from one mouse
myoblast cell nucleus, raising the question of whether the observed
dependence of the binding time on the volume exclusivity and binding
site localization was simply an artifact of the particular cell we
studied. In this work we repeat the computational studies
of~\cite{IsaacsonPNAS2011} within five mouse olfactory sensory neuron
cell nuclei obtained by SXT imaging.  We observe the same qualitative
behavior of the mean binding time on volume exclusivity and binding
site localization as was observed for the median binding time
in~\cite{IsaacsonPNAS2011}. In addition, we now give a theoretical
explanation \emph{why} the mean binding time has this qualitative
behavior.

We begin in the next section by summarizing the SXT imaging data we
use in constructing our mathematical models. It is shown that the
distribution of LACs within each nucleus is bimodal, indicating a
division of the nuclear space into regions with high densities of
material and low densities of material. We infer that the lower mode
corresponds to the most likely regions of euchromatin, less-compact
DNA comprised of the majority of active genes, while the higher mode
corresponds to the most likely regions of heterochromatin,
more-compact DNA thought to contain most silenced
genes~\cite{AlbertsMOLECELLBIO}.

In Section~\ref{S:model} we summarize the mathematical model we
developed in~\cite{IsaacsonPNAS2011}. The protein is assumed to
diffuse in a volume-excluding potential.  Since the underlying SXT
imaging data is a 3D grid of voxels, we assume the protein's motion
can be approximated by a \comment{Markovian} continuous-time random
walk. The volume exclusion potential within a given voxel is chosen
proportional to the normalized LAC of that voxel.  The protein moves
by hopping between neighboring voxels of the 3D grid with jump rates
determined by the protein's diffusion constant and the strength of the
potential difference between the two voxels.
Section~\ref{S:numericalMethod} summarizes the underlying stochastic
simulation algorithm (SSA) we use to simulate the protein's search for
the binding site.

In Section~\ref{S:sims} we repeat the studies
of~\cite{IsaacsonPNAS2011} using volume exclusion potentials
reconstructed from SXT imaging of five cell nuclei. We verify that the
conclusions of~\cite{IsaacsonPNAS2011} still hold in each of the five
nuclei, and demonstrate that for binding sites localized in
euchromatin, \textit{i.e.} regions of low chromatin density, volume
exclusion can lead to decreases in the mean binding time of $23\%$ to
$34\%$ when compared to simulations with no volume excluding potential
(zero volume exclusivity). Finally, in Section~\ref{S:mtTheory} we
develop an analytical theory to explain the observed dependence of the
mean binding time on the volume exclusivity and binding site
localization. The theory suggests that for binding sites in
euchromatin there is an optimal level of volume exclusivity that
balances a reduction in the volume searched in finding the binding
site, with the height of effective potential barriers the protein must
cross during the search process.

\section{Nonuniformity of Nuclear Chromatin Distribution}

To measure the spatial variation in chromatin density within nuclei we
make use of soft X-ray tomographic (SXT) reconstructions of cells. For
an overview of SXT imaging we refer the reader
to~\cite{McDermott:2009ib}.  In this work we use reconstructions of
mouse olfactory sensory neurons, including several mature cells and
one immature cell taken from the data of~\cite{LarabellCell2013}.  The
experimental protocol for obtaining these reconstructions was the same
used in~\cite{LarabellCell2012}.  SXT is similar in concept to medical
X-ray CT imaging, but uses soft X-rays in the ``water window'' which
are absorbed by carbon and nitrogen dense organic matter an order of
magnitude more strongly than by water~\cite{McDermott:2009ib}. As the
absorption process satisfies the Beer-Lambert Law, the measured linear
absorption coefficient (LAC) of one voxel of a 3D reconstruction is
linearly related to the density of organic material within that
voxel~\cite{McDermott:2009ib}. In practice, SXT reconstructions are
able to achieve high resolutions of $50 \, \textrm{nm}$ or less. For
all reconstructions used in this work the underlying voxels were cubes
with sides of length $32 \, \textrm{nm}$.  Another advantage of SXT is
in the minimal preprocessing of cells that is required before imaging.
Cells are cryogenically preserved, but no segmentation, dehydration,
or chemical fixation is necessary. \comment{In
  Appendix~\ref{ap:measErr} we comment on the measurement error of the
  SXT imaging process.}

\begin{figure}
  \centering
  \subfloat[]{    
    \label{fig:cellSXT1} 
    \hspace{-20pt}
    \scalebox{.29}{\includegraphics{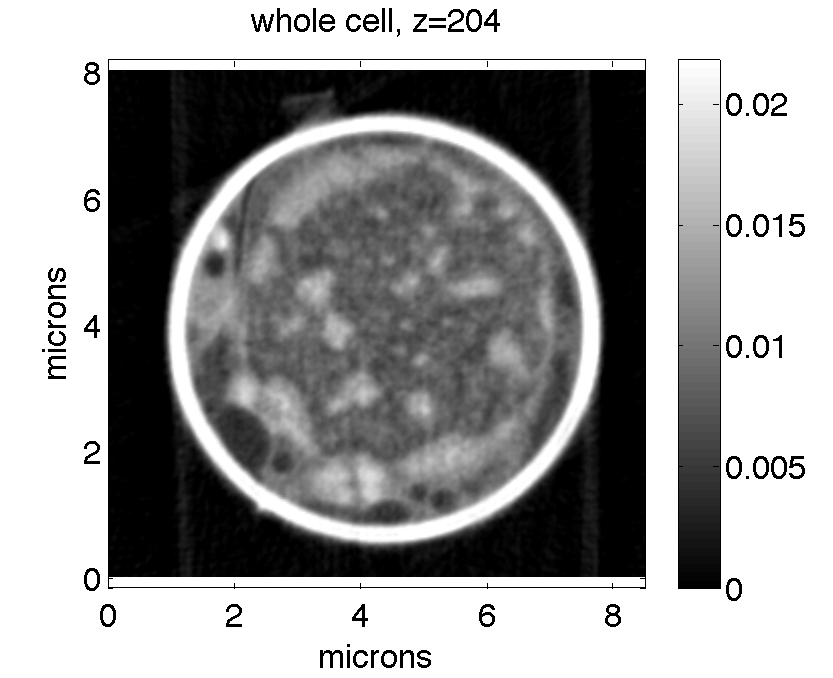}}}
  \subfloat[]{
    \hspace{-5pt}
    \label{fig:cellSXT2} 
    \scalebox{.29}{\includegraphics{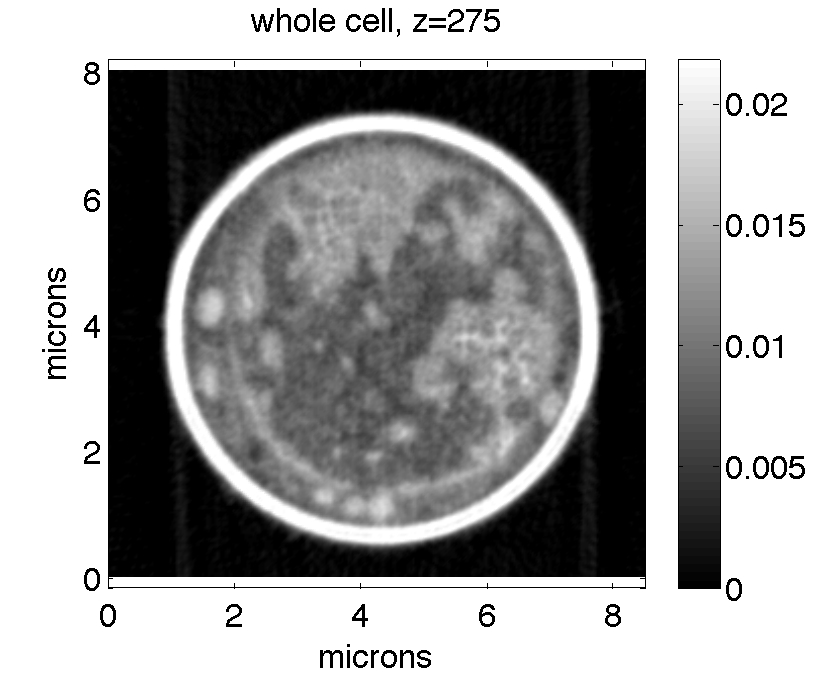}}}\\
  \caption{\small Reconstructed linear absorption coefficients of a
    mature mouse olfactory sensory neuron. The cell was cryogenically
    preserved, and then imaged within a glass capillary (the white
    ring). (a) and (b) show two $z$-plane slices through the
    underlying three-dimensional reconstruction. The cell completely
    fills the capillary in both images. The cell's nucleus is visible
    as the circular structure with large regions of lower LAC values
    (darker pixels). The full 3D reconstruction of this nucleus is
    shown in Figure~\ref{fig:09volRend}. In the figures, LACs have
    units of per voxel. Each LAC value divided by the voxel length of
    $32 \, \textrm{nm}$ would have the more standard units of
    $\textrm{nm}^{-1}$.}
  \label{fig:cellSXT}
\end{figure}
\begin{figure}
  \centering
  \subfloat[]{
    \label{fig:09nuc1} 
    \scalebox{.25}{\includegraphics{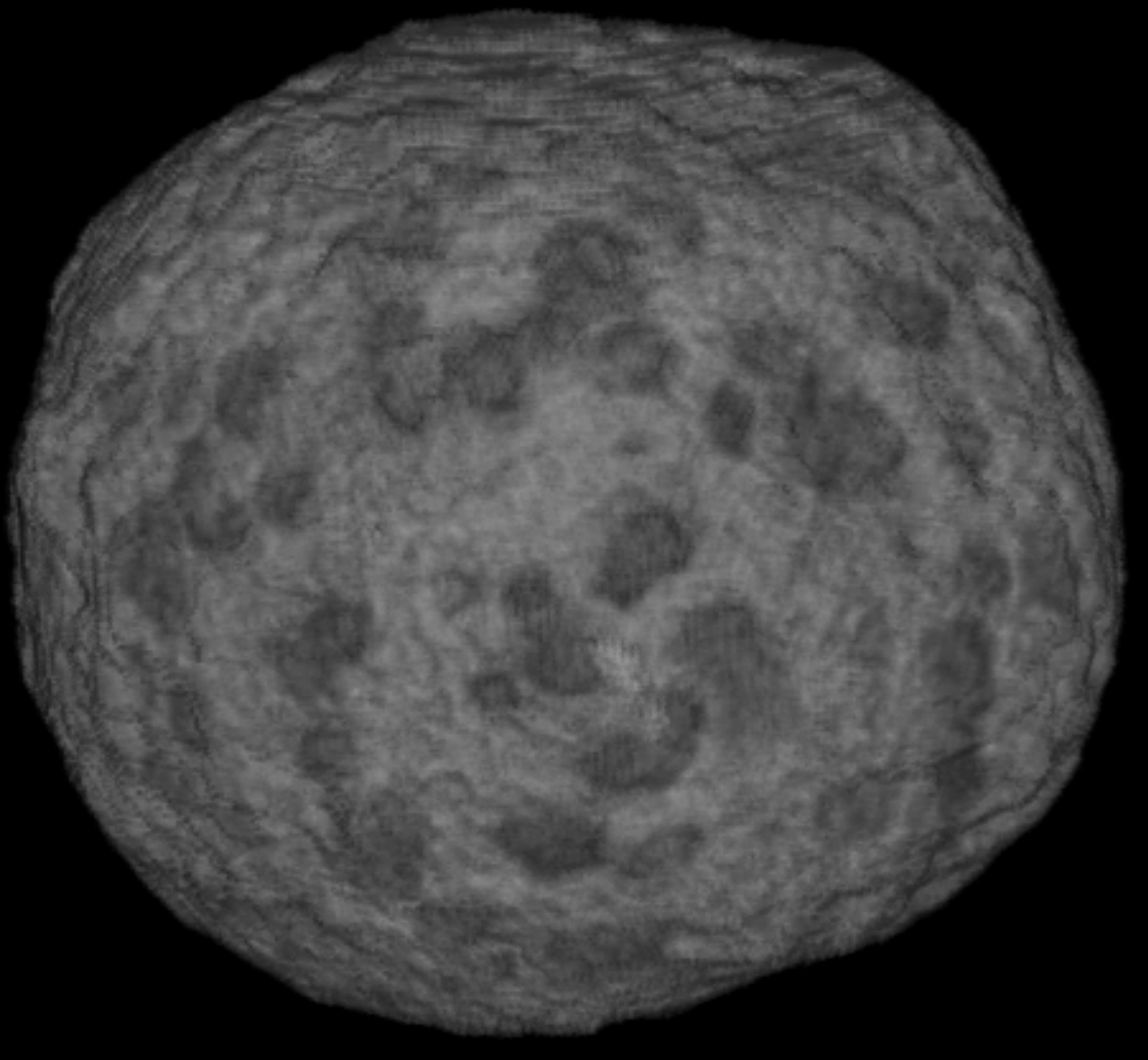}}}
  \subfloat[]{
    \label{fig:09nuc2} 
    \scalebox{.25}{\includegraphics{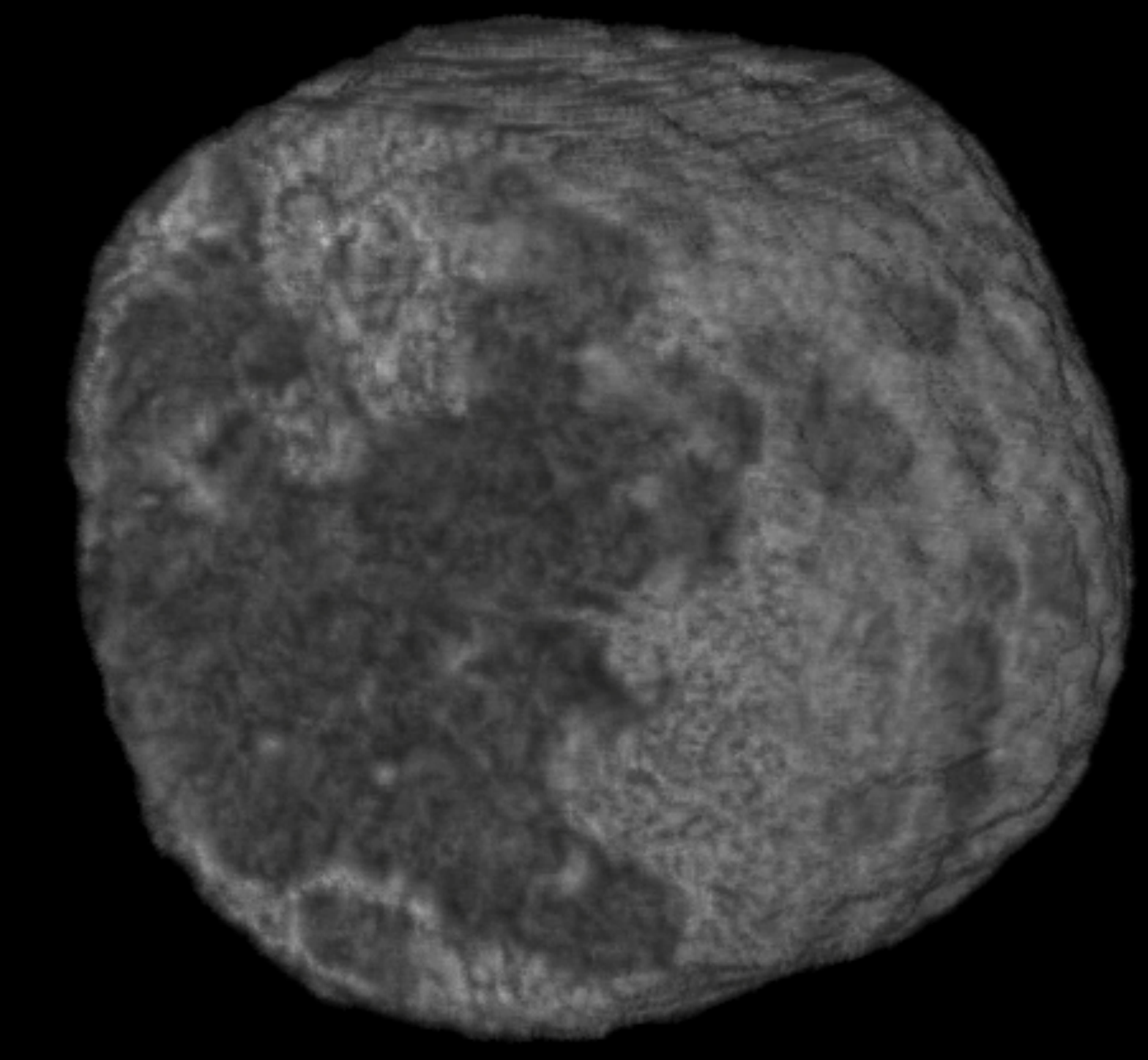}}}
  \subfloat[]{
    \label{fig:09nuc3} 
    \scalebox{.25}{\includegraphics{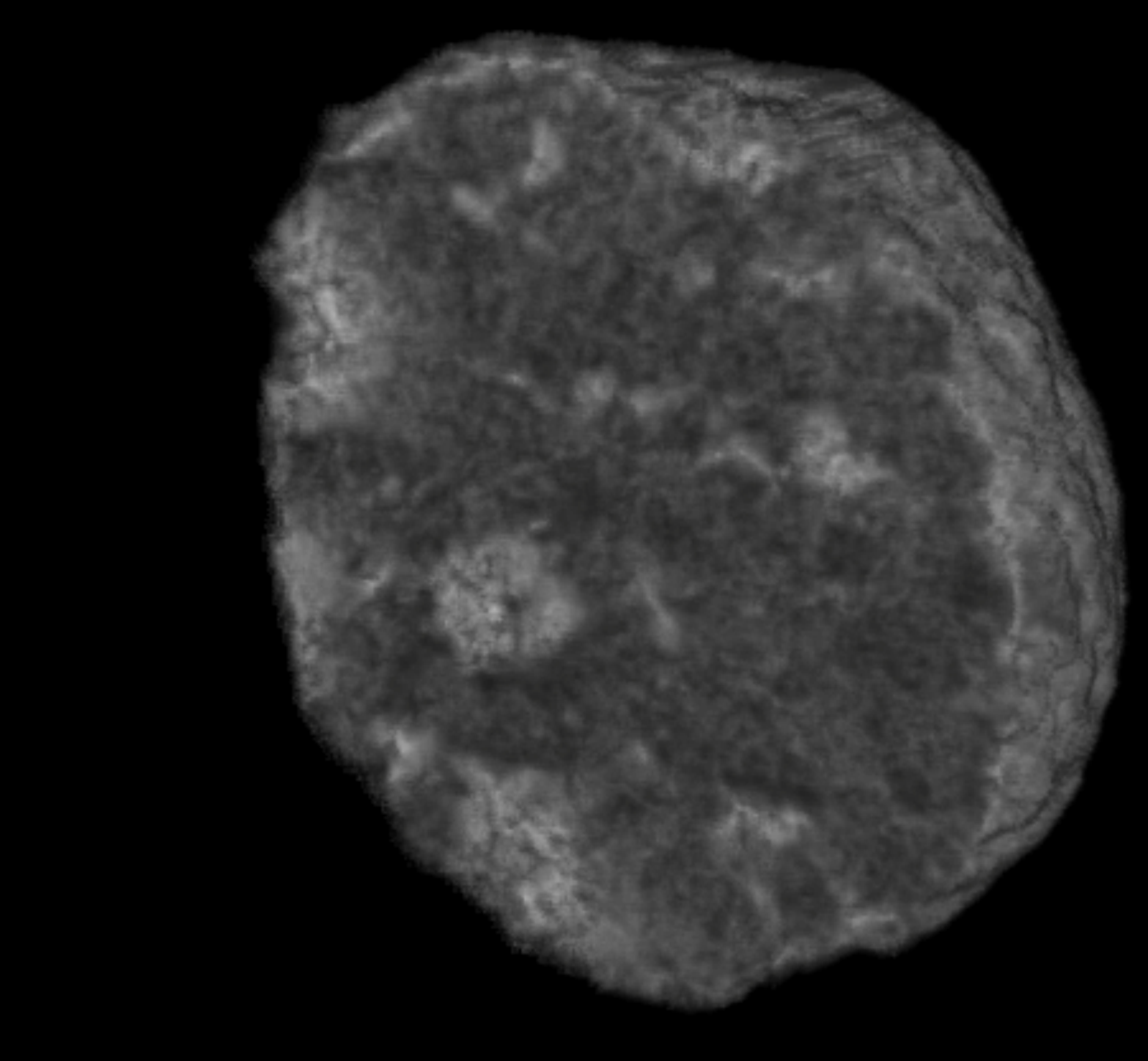}}}
  \subfloat[]{
    \label{fig:09nuc4} 
    \scalebox{.25}{\includegraphics{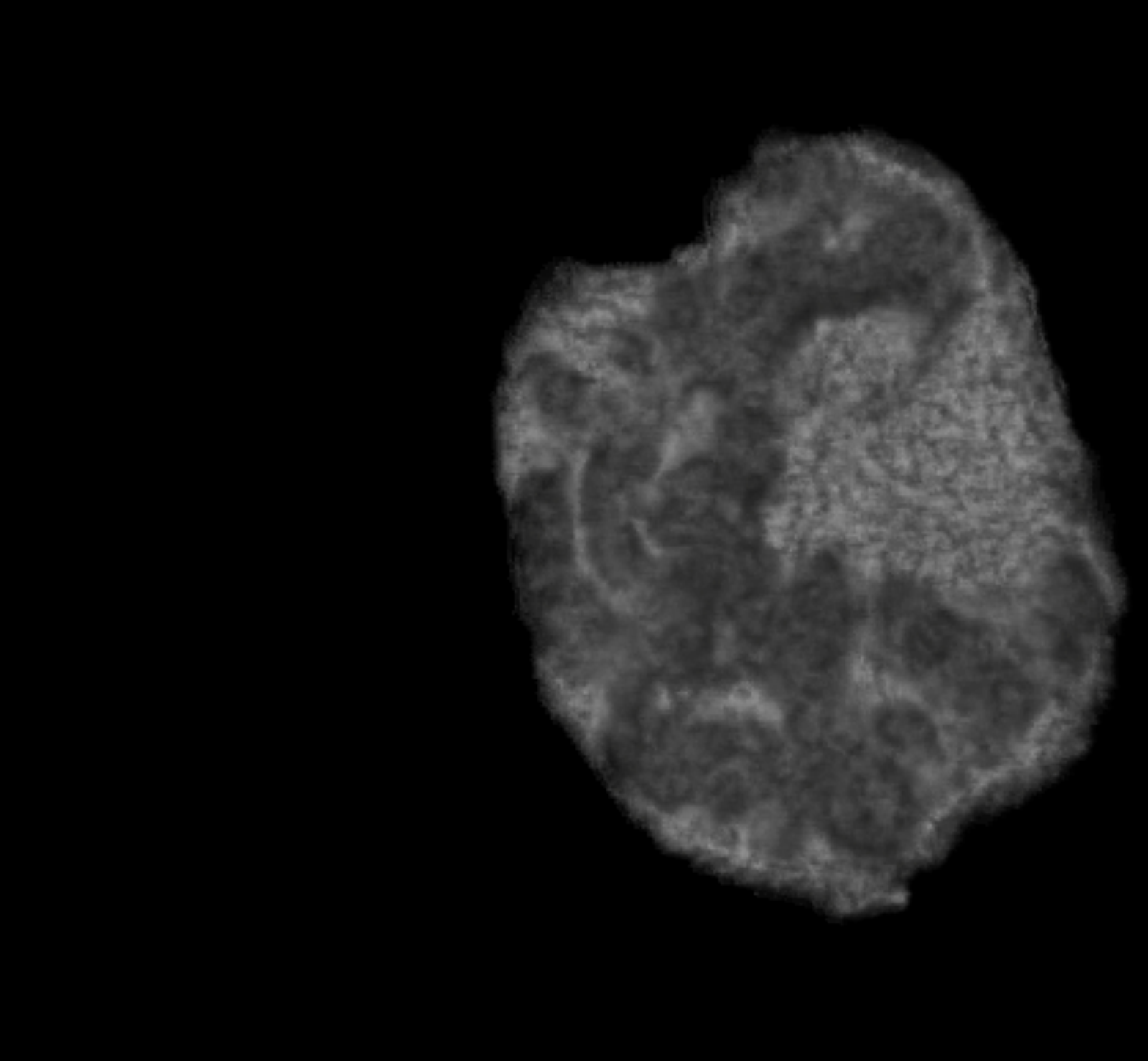}}}
  \caption{\small Nucleus of the cell from Figure~\ref{fig:cellSXT}
    (subsequently labeled by ``09'').  This volume rendering shows the
    LACs of each voxel of the reconstruction. Lighter colors
    correspond to \emph{larger} LACs. (b-d) use clipping planes to
    reveal the interior of the nucleus along the same axis as used for
    the full cell and glass capillary reconstructions shown in
    Figure~\ref{fig:cellSXT}. Movie S1 shows this reconstruction as
    the clipping plane is moved across the nucleus.}
  \label{fig:09volRend}
\end{figure}
As chromatin is the primary organic material within the nuclei of
cells, we subsequently assume the LACs from SXT reconstructions are
directly proportional to the density of chromatin within each voxel of
the reconstruction. For this reason, in the rest of this paper when we
discuss regions of low or high chromatin density we are, more
precisely, discussing regions with low or high densities of organic
material. In Figure~\ref{fig:cellSXT} we show two image plane slices
through a 3D reconstruction of a cryogenically preserved mature mouse
olfactory sensory neuron within a glass capillary. Each image shows
the underlying reconstructed LACs within a given voxel, with smaller
LACs appearing darker (see colorbars for LAC range). In both images
the cell completely fills the capillary, and the cell nucleus is
clearly visible as a circular structure within the cell. The full 3D
reconstruction of the LACs within the nucleus are shown in
Figure~\ref{fig:09volRend}.  Voxels denoting the boundary of the
nucleus were hand traced in Amira~\footnote{Amira, Visualization
  Sciences Group}.  From these traces a mask was produced to label
voxels within the nucleus. In the rest of this paper the nuclear
membrane is assumed to be given by the collection of voxel faces that
are shared by a voxel within and a voxel outside the nucleus.  It is
readily apparent that the nucleus is comprised of regions of low LAC
values interspersed with regions of high LAC values. We subsequently
identify regions of smaller LACs, corresponding to regions of low
density, as euchromatin (regions of less compact DNA, where most
active genes are typically located~\cite{AlbertsMOLECELLBIO}).
Regions of higher density are identified as heterochromatin (more
compact DNA, often containing silenced
genes~\cite{AlbertsMOLECELLBIO}).

\begin{figure}
  \centering
  \subfloat[]{
    \label{fig:09-LAC} 
    \scalebox{.4}{\includegraphics{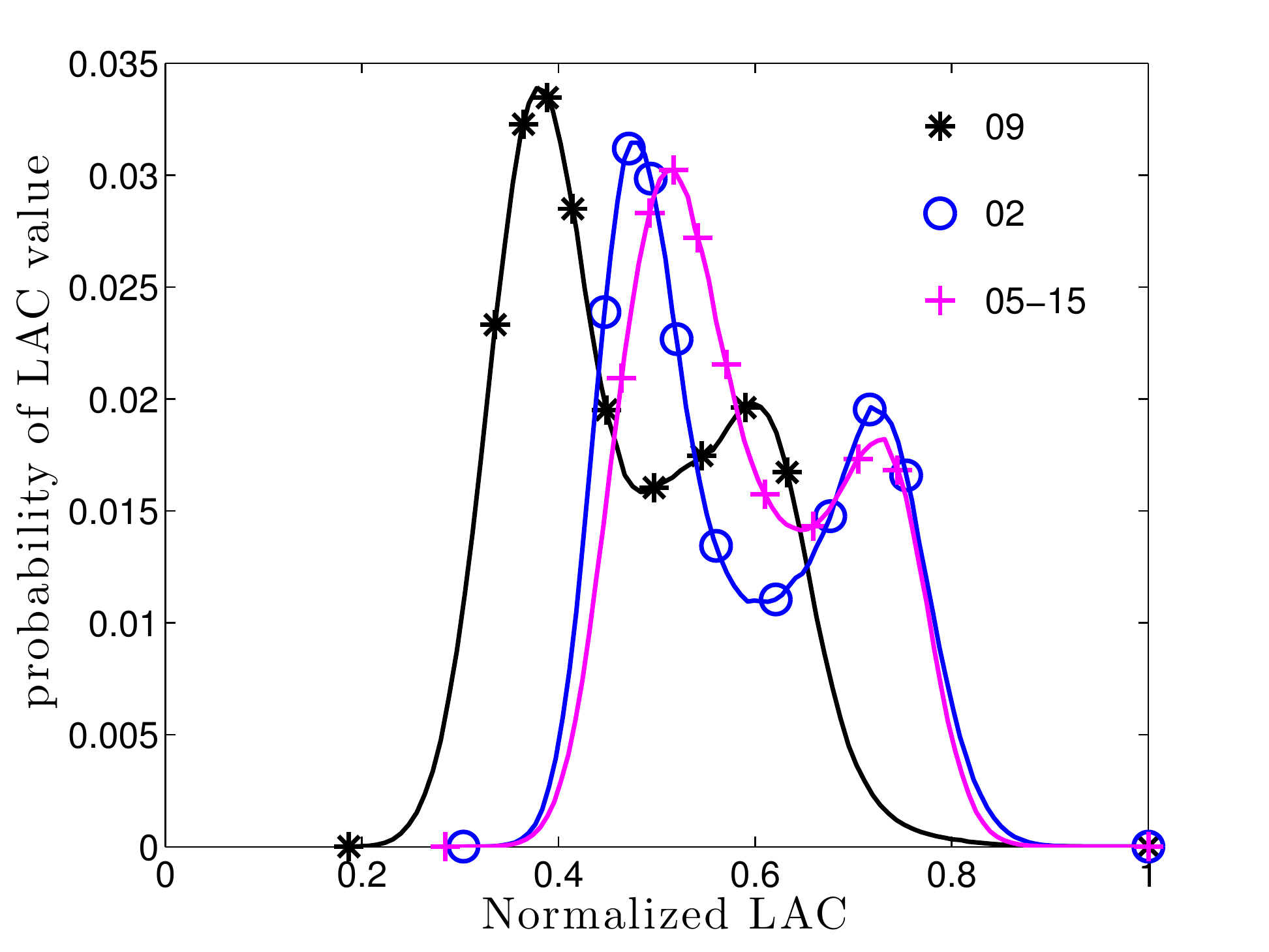}}}
  \subfloat[]{
    \label{fig:05-14-LAC} 
    \scalebox{.4}{\includegraphics{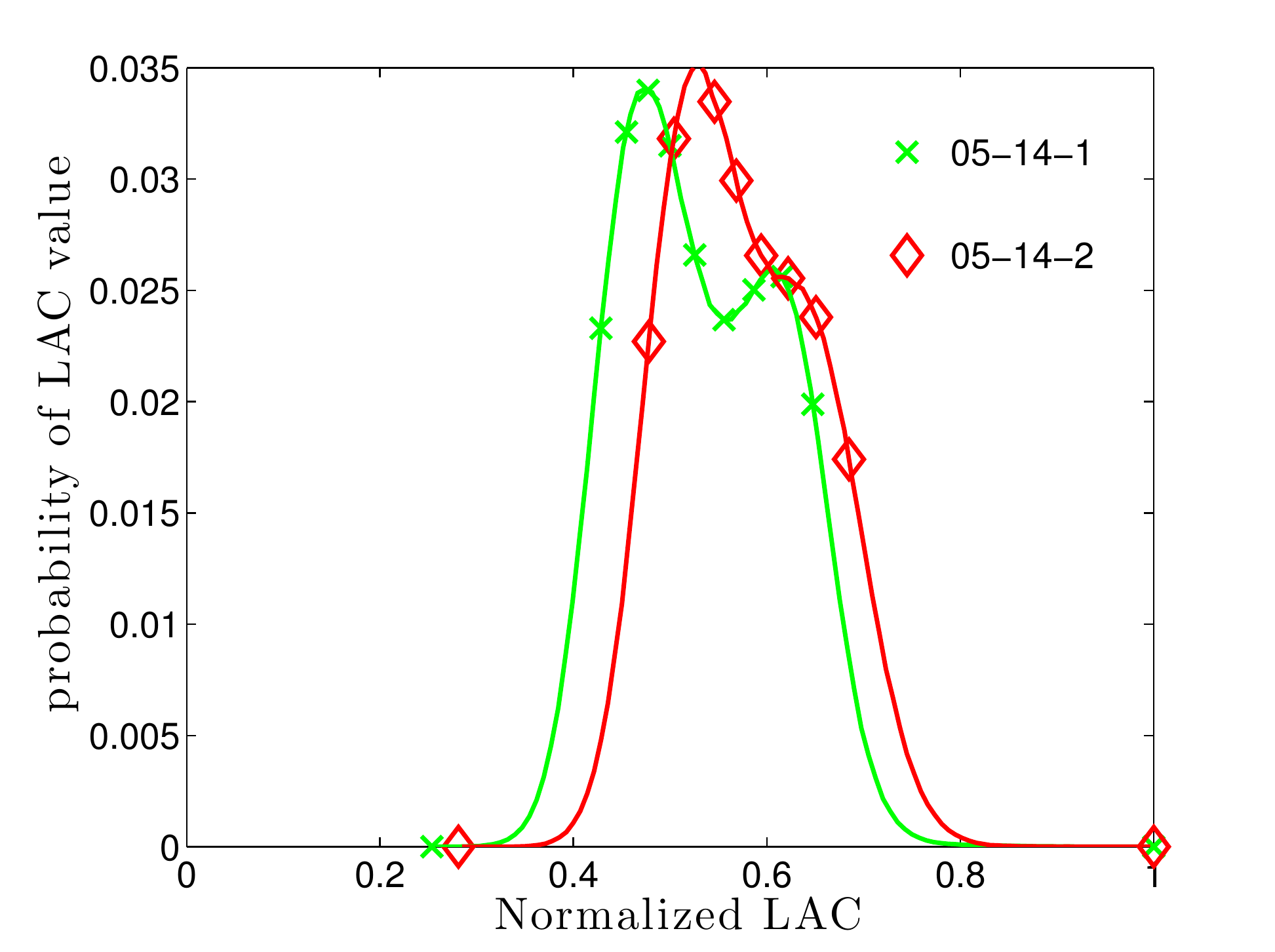}}}
  \caption{\small Histograms of normalized LAC distribution in the
    nuclei of five cells. \comment{Legends give the label subsequently
      used to identify each nucleus.  For each histogram the LAC
      values were normalized by the maximum value within that nucleus.
      Histograms use 100 equally spaced bins between the minimum and
      maximum normalized LAC values. Markers label every tenth
      percentile, beginning with the zeroth. Each nucleus is from
      a mature mouse olfactory sensory neuron, except the
      05-15 dataset which was from an immature mouse olfactory sensory
      neuron. The label for each histogram corresponds to the dataset
      from~\cite{LarabellCell2013} from which that nucleus was
      extracted. The 05-14 dataset contained two cells, with the
      corresponding nuclei labeled by 05-14-1 and 05-14-2.}}
  \label{fig:LACHists}
\end{figure}
Figure~\ref{fig:LACHists} provides further motivation for this
labeling.  There we plot histograms of the normalized LAC distribution
within the nuclei of five cells. (By normalized we mean that the LACs
have been rescaled so that the maximum LAC within each nucleus is
one.) For each nucleus the normalized LAC distribution is bimodal,
illustrating the division of the nucleus into regions of euchromatin
and heterochromatin. Moreover, we see that the first mode of the
distribution occurs between the 20th and 30th percentiles of the
normalized LAC distribution, while the second mode occurs between the
70th and 90th percentiles. We therefore interpret these percentile
ranges as the most likely LACs for euchromatin and heterochromatin
respectively. 
Note, the bimodal LAC distribution observed in each nucleus is
fundamentally different than the unimodal DAPI stain fluorescence
distribution we saw in~\cite{IsaacsonPNAS2011} (which was peaked at
zero intensity, see Figure~1C of~\cite{IsaacsonPNAS2011}). 

\section{Mathematical Model} \label{S:model} We are interested in
studying the statistics of the time required for a diffusing
regulatory protein to find specific binding sites within the nucleus
of a eukaryotic cell. The mathematical model and biological
assumptions we describe in this section are based on the model we
developed in~\cite{IsaacsonPNAS2011}. For completeness we summarize
them here, but refer the reader to~\cite{IsaacsonPNAS2011} for further
detail.

We assume that in the absence of chromatin, inside the nucleus the
protein would move by Brownian motion with a fixed diffusivity.
Volume exclusion caused by the varying spatial density of chromatin is
modeled by a repulsive potential that imparts drift to the protein's
movement. The strength of the potential is assumed to be proportional
to the density of chromatin at a given point in space.  Regions of
higher chromatin density will therefore be more volume excluding,
\textit{i.e.}  more difficult for proteins to enter, than regions
of low density. Note, there are many other possible interactions that
may influence the protein's search process.  These include
non-specific DNA binding interactions, trapping caused by such
interactions or local chromatin structure, and the possibility of
protein diffusion along
DNA~\cite{HalfordBindRatesAreDiffLim09,ElfSingMolSliding07}. The model
we describe focuses on how the search process for specific binding
sites is influenced by the assumption that regions of high DNA density
are more difficult to enter.

As described in the previous section, the measured soft X-ray
tomography (SXT) linear absorption coefficients (LACs) are directly
proportional to the density of organic matter within a given voxel. We
 assume here that the LAC gives a direct measure of the density of
chromatin in a voxel. Let $I \subset \Z^3$ denote the set of voxels
from a SXT reconstruction that comprise the nucleus of a cell.  For
$\bi \in I$ a given voxel within the nucleus, we denote by
$\ell_{\bi}$ the normalized LAC for that voxel.  By normalized we mean
that the $\ell$ has been rescaled so that $\max_{\bi \in I} \ell_{\bi}
= 1$. We assume the potential of the $\bi$th voxel is given by
\begin{equation} \label{eq:potentDef}
  \phi_{\bi} = \phimax \ell_{\bi},
\end{equation}
where the potential maximum $\phimax$, subsequently called the
\emph{volume exclusivity}, is a parameter of our model. When $\phimax
= 0$ the particle will simply diffuse, while as $\phimax \to \infty$
it will become increasingly difficult for the protein to enter
regions of high chromatin density.

As the imaging data are defined by a mesh of voxels, we approximate
the diffusion of the protein within the nucleus as a
\comment{Markovian} continuous-time random walk among these voxels.
The protein moves by hopping between neighboring voxels. The
probability per unit time that the protein hops from one voxel to a
neighbor is determined by the diffusion constant of the protein, the
edge length of the voxels, and the strength of the potential barrier
crossed in hopping from one voxel to another. Let $P_{\bi}(t)$ label
the probability the protein is in the $\bi$th voxel at time $t$. In
what follows we assume the DNA binding site is given by one specific
voxel of the mesh, $\bib$, and that upon reaching this voxel the
protein immediately binds. Using this condition, our model can be
interpreted as an approximation of the search process by the protein
for a small region containing the binding site (\textit{i.e.}  the
voxel $\bib$). The binding reaction then gives the reactive boundary
condition that
\begin{equation} \label{eq:rdmeBC}
  P_{\bib}(t) = 0.
\end{equation}
\comment{Note, this choice of boundary condition assumes the protein
  is immediately removed from the system upon hopping into the voxel
  containing the binding site. This is equivalent, insofar as the rest
  of the system is concerned, to the assumption that the protein stays
  at the binding site once it has arrived there. It is purely a matter
  of bookkeeping whether we say that the protein is removed from the
  system upon arrival at the binding site or remains at the binding
  site once it has arrived there.}

Let $\alpha_{\bi \bj}$ label the probability per unit time the protein
will hop to voxel $\bi$ when within voxel $\bj$. The master equation
for the probability the protein is in the $\bi$th voxel at time $t$ is
then the linear system of ODEs
\begin{equation} \label{eq:rdme}
  \D{P_{\bi}}{t}(t) = \sum_{\bj \in I} \alpha_{\bi \bj} P_{\bj}(t) - \alpha_{\bj \bi} P_{\bi}(t), 
  \quad \forall \bi \in I - \{\bib\},
\end{equation}
where $I - \{\bib\}$ denotes the set of voxels, $I$, with the $\bib$
voxel removed.  Note this model implicitly enforces the no-flux
boundary condition that the protein can not leave the nucleus.

The spatial hopping rates $\alpha_{\bi \bj}$ were derived in the
supplement to~\cite{IsaacsonPNAS2011}. They are chosen so that in the
absence of the reactive boundary condition~\eqref{eq:rdmeBC}, the
master equation~\eqref{eq:rdme} would converge as the voxel size
approaches zero to a Fokker-Planck partial differential equation (PDE)
describing the drift-diffusion of the protein. Let $D$ the diffusion
constant of the protein, $h$ the edge length of each voxel, $\kb$
Boltzmann's constant, and $T$ the temperature of the nucleus.
\comment{We found in~\cite{IsaacsonPNAS2011} that when $\bi$ and $\bj$
  are nearest neighbor voxels along a coordinate axis the choice
\begin{equation} \label{eq:rdmeRates}
  \alpha_{\bi \bj} = 
  \frac{2 D}{h^2} \frac{1}{\exp((\phi_{\bi}-\phi_{\bj})/\kb T) + 1},
\end{equation}
with $\alpha_{\bi \bj}=0$ for all other voxel pairs,} provides a
second order spatial discretization of the corresponding Fokker-Planck
PDE. As the potential barrier to hop from voxel $\bj$ to $\bi$ grows
($\phi_{\bi} - \phi_{\bj} \to \infty$), we see that $\alpha_{\bi \bj}
\to 0$. As the potential becomes constant ($\phi_{\bi} - \phi_{\bj}
\to 0$), we recover a discretization of the standard discrete
Laplacian with the corresponding jump rates
\begin{equation*}
  \alpha_{\bi \bj} = \frac{D}{h^2}.
\end{equation*}
Note the not-quite-obvious consequence of~\eqref{eq:rdmeRates} that
\begin{equation*}
  \frac{\alpha_{\bi \bj}}{\alpha_{\bj\bi}} = e^{-\paren{\phi_{\bi} - \phi_{\bj}} / \kb T}.
\end{equation*}
This ensures that the steady state solution to~\eqref{eq:rdme}, in the
absence of the reactive boundary condition~\eqref{eq:rdmeBC}, is the
discrete Gibbs-Boltzmann distribution
\begin{equation} \label{eq:rdmeSS}
  \Pss_{\bi} = \frac{e^{-\phi_{\bi} / \kb T}}{\sum_{\bj \in I} e^{-\phi_{\bj} / \kb T}}.
\end{equation}
This steady-state will be used in Section~\ref{S:mtTheory} to
estimate the explicit dependence on $\phimax$ of the mean time for
the protein to find the binding site, $\bib$.

We assume that the protein begins its search for the binding site from
a nuclear pore within the nuclear membrane.  As nuclear pores are not
explicitly visible by soft X-ray tomography, we make the approximation
that each voxel on the border of the nucleus contains the same
fraction of the total number of nuclear pores. Instead of studying the
protein's search from one specific pore (voxel), \comment{at the start
  of \emph{each} individual simulation} we choose the protein's
initial position from a uniform distribution among all voxels on the
border of the nucleus.  The corresponding initial condition for the
master equation~\eqref{eq:rdme} is therefore
\begin{equation} \label{eq:rdmeIC}
  P_{\bi}(0) =
  \begin{cases}
    \frac{1}{N_{\textrm{bnd}}}, \quad \bi \in I_{\textrm{bnd}}, \\
    0, \quad \bi \notin I_{\textrm{bnd}},
  \end{cases}
\end{equation}
where $I_{\textrm{bnd}}$ labels the set of voxels on the boundary of
the nucleus and $N_{\textrm{bnd}}$ denotes the number of voxels in
this set.

We shall also investigate an extension of the preceding model where
$\bib$ is allowed to be a random variable. To understand the
difference in binding times for binding sites localized in euchromatin
versus heterochromatin, we allow $\bib$ to be chosen from a uniform
distribution among all voxels with LACs between two specified
percentiles of the nuclear LAC distribution. Binding sites placed in
euchromatin were chosen from low percentiles, usually the 20th to
30th, near the first mode of the LAC distribution (see
Figure~\ref{fig:LACHists}). \comment{For the fluorescence imaging data
  we used in~\cite{IsaacsonPNAS2011}, this percentile range
  corresponded to the first in which there was a non-zero fluorescence
  level. For the SXT data this range represents the most likely LAC
  values for euchromatin containing voxels. (It is unknown if voxels
  with sufficiently low LAC values actually contain chromatin.
  Similarly, it has yet to be determined experimentally if there is a
  precise LAC value above which voxels may be assumed to contain
  heterochromatin.)} To model heterochromatin we used higher
percentiles near the second mode, such as the 70th to 80th.

\section{Numerical Solution Method} \label{S:numericalMethod} For a
specified value of $\bib$, the master equation~\eqref{eq:rdme} with
boundary condition~\eqref{eq:rdmeBC} and initial
condition~\eqref{eq:rdmeIC} is a linear system of ODEs. A typical
reconstruction of a nucleus contains on the order of 2.7 million
voxels (slightly more than would be contained in a 128 by 128 by 128
Cartesian mesh). While such a system of ODEs can be solved directly,
allowing $\bib$ to be a random variable would potentially require the
system to be solved many times to obtain good statistical estimates of
the binding times.  For this reason we simulated the underlying
continuous-time random walk of the protein between the voxels instead
of directly solving~\eqref{eq:rdme}. In these simulations the protein
hops from voxel $\bi$ to neighbor $\bj$ with probability per unit time
$\alpha_{\bj \bi}$.  \comment{When the protein hops into the voxel
  containing the binding site}, $\bib$, the simulation is terminated
and the time the protein entered the voxel recorded.  Exact
realizations of this stochastic process can be generated by the
stochastic simulation algorithm (SSA), also known as the Gillespie
method or Kinetic Monte
Carlo~\cite{GillespieJPCHEM1977,KalosKMC75,GibsonBruckJPCHEM2002}.

Our numerical simulation algorithm can be summarized as follows
\begin{enumerate}[noitemsep]
\item Pre-calculate the jump rates, $\alpha_{\bi \bj}$.
\item Choose the binding site location, $\bib$. This is either
  specified, or sampled from voxels within specified percentiles of
  the LAC distribution (\textit{i.e.} euchromatin or heterochromatin
  regions).
\item Sample the initial position of the protein from $P_{\bi}(0)$. 
\item Use the SSA to simulate the motion of the protein until the
  time, $\tau$, that it hops into the voxel $\bib$.
\item Repeat from step 2 until the desired number of simulations have
  been run.
\end{enumerate}

\section{Mean Time to Find a DNA Binding Site} \label{S:sims}
We now investigate the statistics of the first passage time, $\tau$, for
the protein to find the binding site, and how these statistics depend
on the binding site position and volume exclusivity, $\phimax$. We
focus on the survival probability that the protein has not found the
binding site by time $t$,
\begin{equation*}
  \prob \brac{\tau > t} = \sum_{\bi \in I} P_{\bi}(t),
\end{equation*}
and the mean binding time
\begin{equation*}
  \avg{\tau} = \int_0^{\infty} \prob \brac{\tau > t} \, dt. 
\end{equation*}
For all reported simulations, $\prob \brac{\tau > t}$ and associated
$95\%$ confidence intervals were estimated using MATLAB's
\texttt{ecdf} routine. \comment{$\avg{\tau}$ and associated confidence
  intervals were estimated using MATLAB's \texttt{mean} routine, while
  the standard error was estimated using standard deviations
  determined by MATLAB's \texttt{std} routine.}

In the following we choose $D = 10 \, \mu \textrm{m}^2
\textrm{s}^{-1}$, and report $\phimax$ in units of $\kb T$. For each
SXT reconstruction the voxels were cubic with edge length $32 \,
\textrm{nm}$. In all simulations spatial units were in $\mu \text{m}$,
so that $h = .032 \, \mu \text{m}$, and time had units of seconds.

\begin{figure}[t]
  \centering
  \subfloat[]{
    \label{fig:fixedLocMean} 
    \scalebox{.4}{\includegraphics{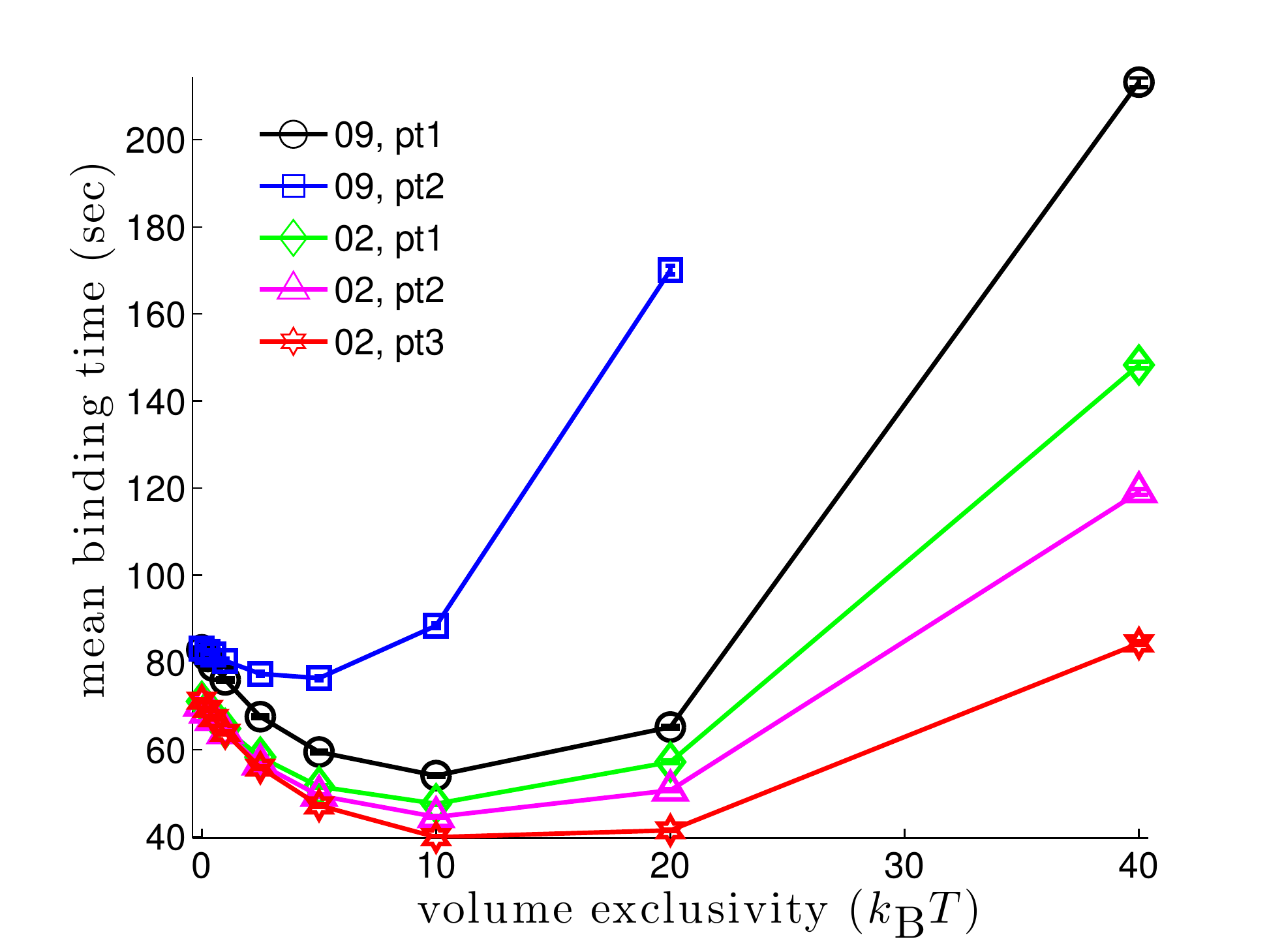}}}
  \subfloat[]{
    \label{fig:fixedLocCDF}  
    \scalebox{.4}{\includegraphics{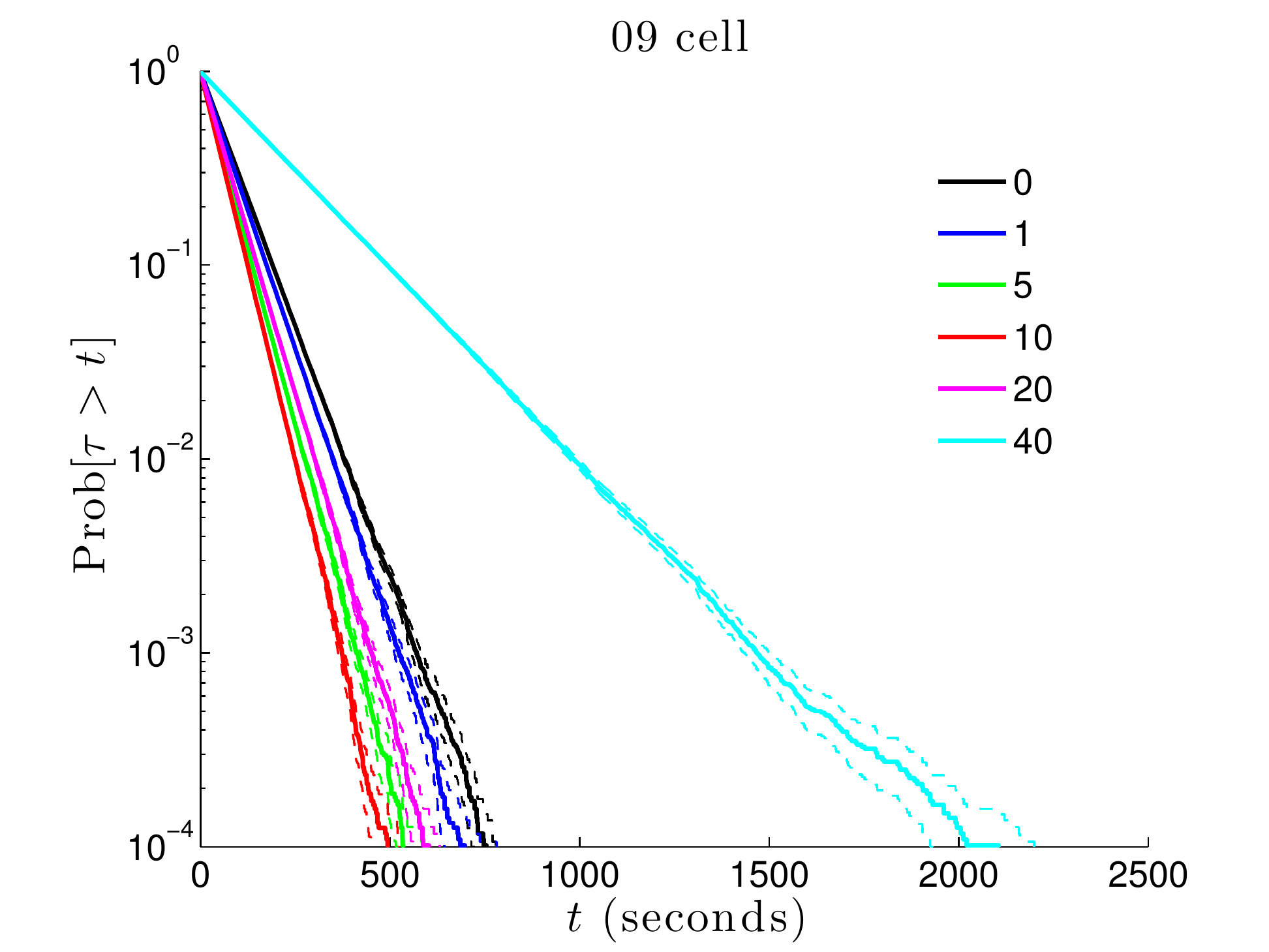}}}    
  \caption{\small Mean binding time and survival time distribution for
    binding sites at fixed locations in euchromatin.  (a) Mean binding
    time as the volume exclusivity, $\phimax$, is varied.  Inset gives
    the cell nucleus and target point label for each curve. Each data
    point was obtained from 128000 simulations.  $95\%$ confidence
    intervals are shown for each point, but often smaller than the
    marker size. As $\phimax$ is increased we see each curve initially
    approaches a minimum before diverging to $\infty$.  (b) Survival
    time distribution for the ``pt1'' binding site from the 09 cell
    nucleus.  The inset gives the volume exclusivity, $\phimax$, in
    units of $\kb T$.  Each curve was generated from binding time
    statistics for 128000 simulations. The dashed lines about each
    solid curve give $95\%$ confidence intervals.  Linearity of $\prob
    \brac{\tau > t}$ with the logarithmic $y$-axis suggests the binding
    time is approximately exponential. }
  \label{fig:fixedLoc}
\end{figure}
We first examined the time to find a fixed binding site within regions
of euchromatin. Several voxels within the euchromatin regions of the
09 and 02 cell nuclei were sampled randomly from all voxels in the
20th to 30th percentiles of the nuclear LAC distribution (see
Figure~\ref{fig:LACHists}). For each fixed binding site 128000
simulations were run, and the time that the protein first encountered
the binding site was recorded. In Figure~\ref{fig:fixedLoc} we examine
the statistics of the binding time, $\tau$, for five different binding
sites within two cell nuclei (labeled by ``09, pt1'', ``09, pt2'',
``02, pt1'', ``02, pt2'', and ``02, pt3'').
Figure~\ref{fig:fixedLocMean} shows the mean binding time as a
function of $\phimax$ for each of the five binding sites. We see that
as the volume exclusivity is increased from zero the mean binding time
decreases to a minimum. As the volume exclusivity is further increased
the mean binding time then increases dramatically. This behavior can
also be seen in Figure~\ref{fig:fixedLocCDF}, where we show the
survival time distribution for the binding site ``pt1'' from the 09
cell nucleus as the volume exclusivity, $\phimax$, is increased.
Notice that each curve is linear with a logarithmic $y$-axis,
suggesting that the time to find the binding site is approximately an
exponential random variable.

\begin{figure}
  \centering
  \subfloat[]{
    \label{fig:randLocMean09} 
    \scalebox{.4}{\includegraphics{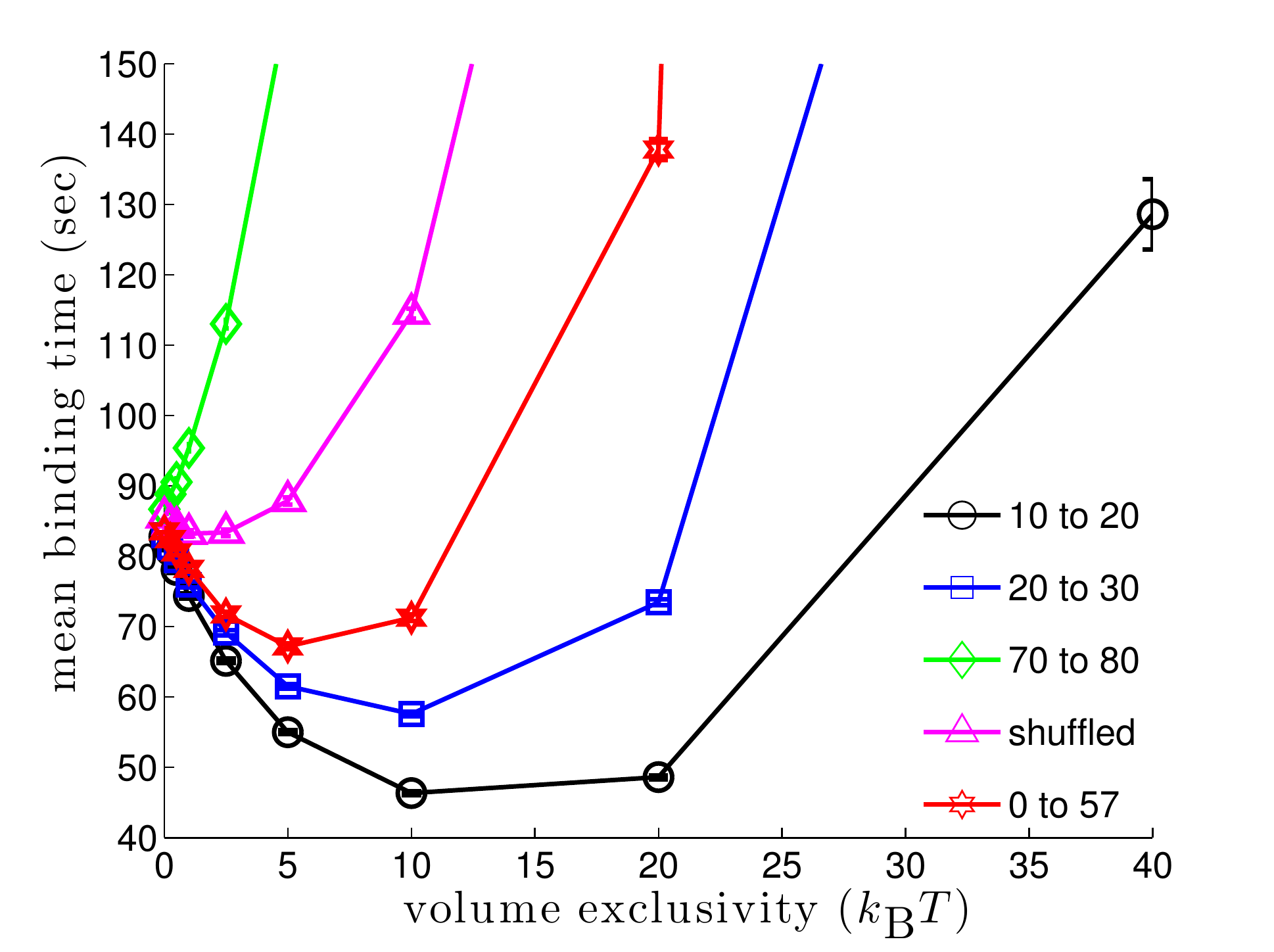}}}
  \subfloat[]{
    \label{fig:randLocMeanMultcells} 
    \scalebox{.4}{\includegraphics{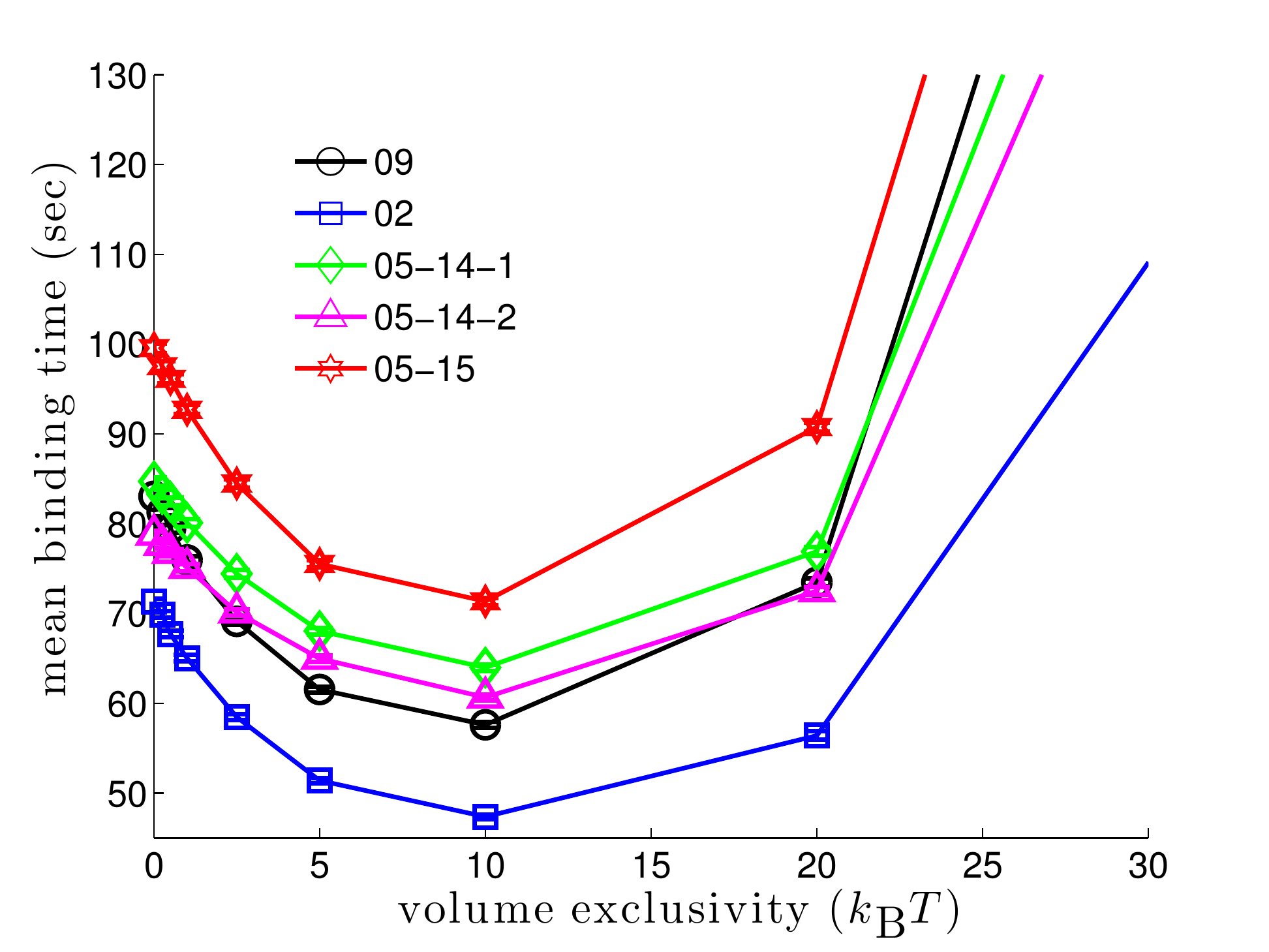}}}
  \caption{\small Mean binding time for binding sites randomly
    localized in subsets of the nucleus as the volume exclusivity is
    varied. For each simulation the target binding site was sampled
    from a uniform distribution among all voxels within fixed
    percentile ranges of the nuclear LAC distribution. Each data point
    was calculated from 128000 simulations. (a) Mean binding times in
    the 09 cell.  The inset gives the percentile range used for each
    curve.  The black (10 to 20) and blue (20 to 30) curves localized
    binding sites in regions of euchromatin, see
    Figure~\ref{fig:09-LAC}. \comment{The red (0 to 57) curve
      localized binding sites anywhere before the LAC value
      giving the minimum between the two modes of the LAC
      distribution.}  The green (70 to 80) curve localized binding
    sites in heterochromatin.  For the magenta curve the values of the
    LAC distribution were randomly shuffled among the voxels of the
    nucleus. Binding sites were then chosen from a uniform
    distribution among voxels with LAC values in the 20th to 30th
    percentiles of the LAC distribution.  For binding sites in
    heterochromatin, or when using the shuffled potential, the mean
    binding time diverges as the volume exclusivity is increased. (b)
    Mean binding times for binding sites randomly chosen at the start
    of each simulation within the 20th to 30th percentile of the LAC
    distribution for five different cell nuclei.  For each cell,
    binding sites in euchromatin were found fastest when $\phimax
    \approx 10 \, \kb T$.}
  \label{fig:randLocMeans}
\end{figure}
\begin{figure}
  \centering
  \subfloat[]{
    \label{fig:randLocCDF09}  
    \scalebox{.4}{\includegraphics{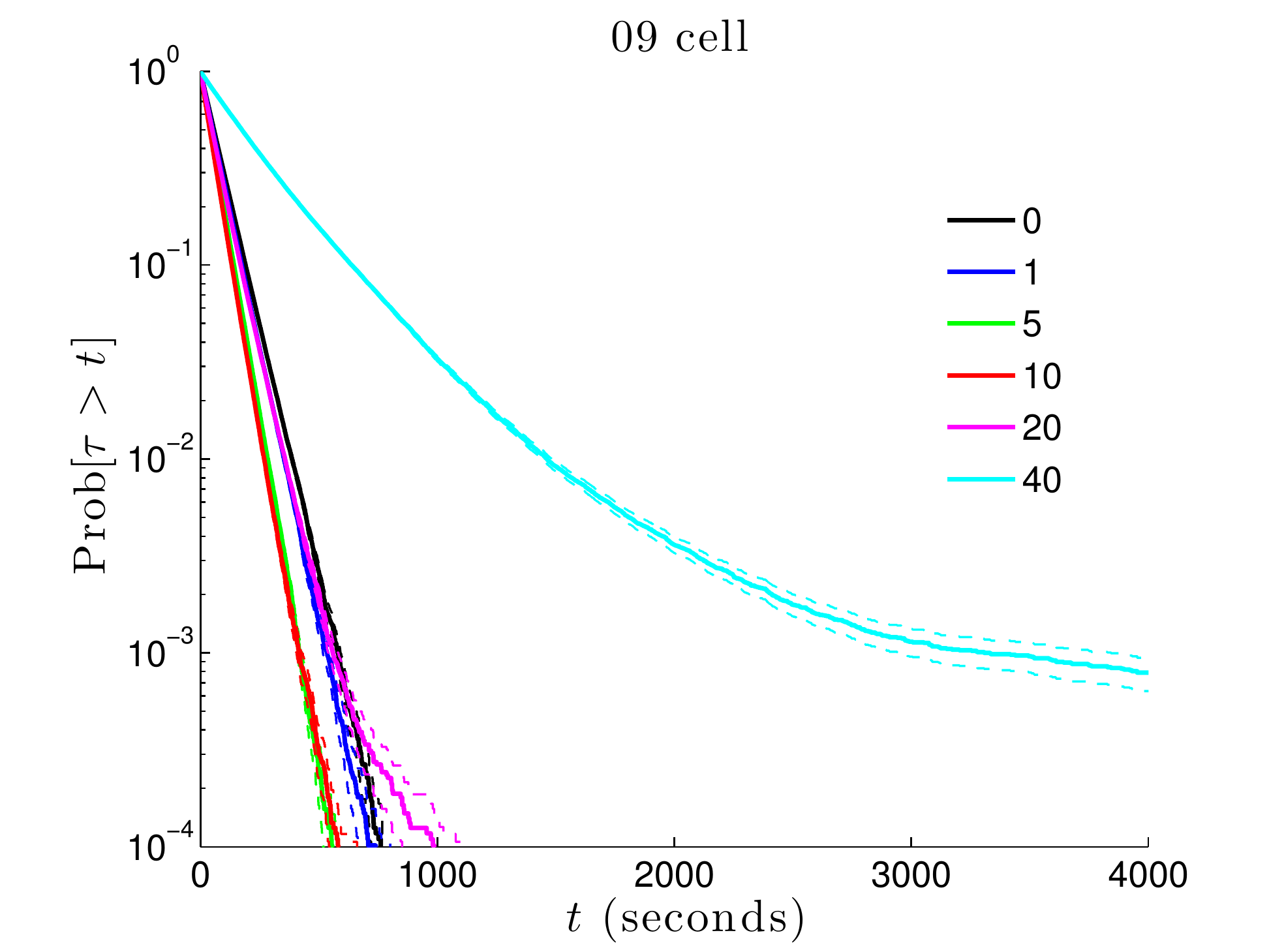}}}    
  \subfloat[]{
    \label{fig:randLocCDF09Zoom}  
    \scalebox{.4}{\includegraphics{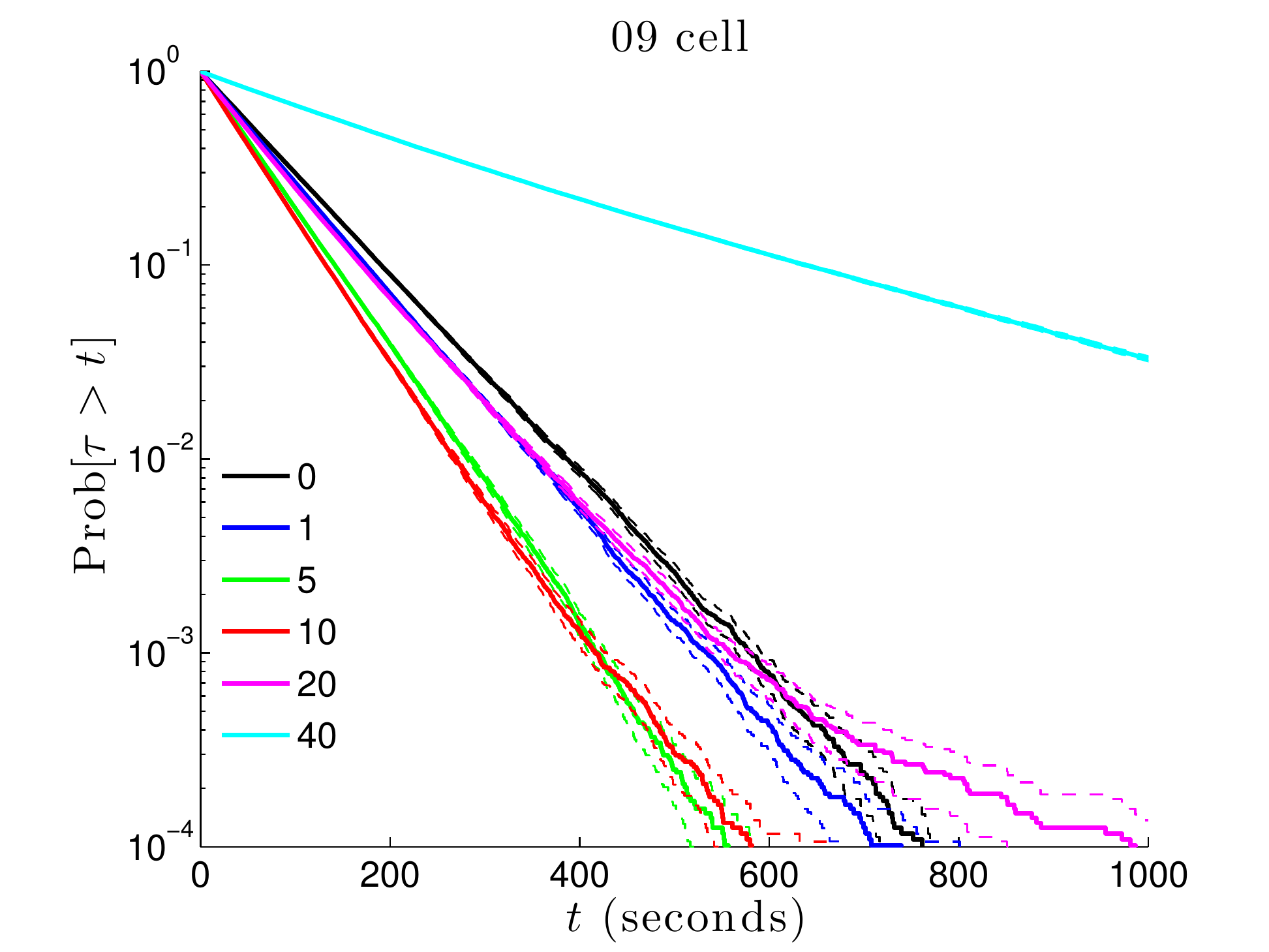}}}    
  \caption{\small Survival time distributions for binding sites
    randomly localized in euchromatin. For each simulation one voxel
    in euchromatin (the 20th to 30th percentile of the LAC
    distribution) is randomly chosen to represent the binding site.
    Each curve is then estimated from 128000 such simulations. Insets
    give the volume exclusivity, $\phimax$, for every curve. (a) For
    large values of the volume exclusivity the survival probability is
    no longer well-approximated by an exponential distribution for all
    times. (b) Reducing the scale of the $t$-axis we can see that as
    $\phimax$ is increased the survival distribution initially shits
    to the left, leading to faster binding times. As $\phimax \to
    \infty$ the distribution shifts back to the right.}
  \label{fig:randLocCDFs}
\end{figure}

Figures~\ref{fig:randLocMeans} and~\ref{fig:randLocCDFs} show how the
binding time varies when the binding site is localized in different
subregions of the nucleus. For each value of $\phimax$, a percentile
range of the LAC distribution was specified and 128000 simulations
were run.  At the beginning of each simulation the binding site was
sampled from a uniform distribution among all voxels within the
percentile range. As such, the figures illustrate the average of the
mean binding time and survival probability distribution over binding
sites localized in different percentile ranges of the LAC
distribution.  Figure~\ref{fig:randLocMean09} illustrates how the mean
binding time varies in the 09 nucleus as a function of $\phimax$ when
binding sites were localized in euchromatin (the 10th to 20th and 20th
to 30th percentile ranges) versus heterochromatin (the 70th to 80th
percentile range). When the binding site was localized in regions of
higher chromatin density the mean binding time increased. Moreover,
while localizing the binding site within euchromatin leads to a
minimum in the mean binding time for non-zero volume exclusivity
(about $10 \, \kb T$), this behavior is lost for binding sites
localized in heterochromatin.  For the latter the mean binding time
simply increases as $\phimax$ increases.
Figure~\ref{fig:randLocMeanMultcells} illustrates that the appearance
of a minimum for non-zero volume exclusivity when binding sites are in
regions of euchromatin (20th to 30th percentile range) persists across
all five of the nuclei we studied.  Moreover, for each nucleus the
minimum appears at $\phimax \approx 10 \, \kb T$.

\comment{The preceding simulations focused on binding sites localized
  near the most likely regions of euchromatin and heterochromatin.  We
  also examined the behavior of the binding time for binding sites
  localized in broader regions. The zeroth to 57th percentile curve in
  Figure~\ref{fig:randLocMean09} corresponds to localizing the binding
  site anywhere before the minimum located between the two modes of
  the LAC distribution of the 09 cell, see Figure~\ref{fig:09-LAC}.
  This minimum was defined to be the center of the bin between the two
  modes of the 09 nucleus histogram in Figure~\ref{fig:09-LAC} that
  had the smallest height, approximately the 57th percentile of the
  LAC distribution.  In~\cite{LarabellCell2013}, the transition point
  between euchromatin and heterochromatin was assumed to be given by
  averaging the LAC values at the two modes. For the 09 nucleus, this
  transition point differs from the location of the minimum between
  the two modes by less than $.1\%$.  Figure~\ref{fig:randLocMean09}
  shows that within this broader region a minimal mean binding time
  still occurs for a nonzero value of the volume exclusivity, but the
  overall decrease in the mean binding time relative to that when the
  volume exclusivity is zero is smaller than for binding sites
  localized near the mode (20th to 30th percentile).  In particular,
  for sites sampled from the zeroth to 57th percentile range we
  observe that the smallest mean binding time is $20\%$ faster than
  that when the volume exclusivity is zero, while for sites sampled
  from the 20th to 30th percentile range the smallest mean binding
  time is $31\%$ faster. This decrease in the observed speed-up of the
  mean binding time arises from the substantially longer time needed
  to find binding sites within voxels having the highest LAC values.
  The smallest median binding time, which is less sensitive to
  outliers, is $27\%$ faster than the median binding time with zero
  volume exclusivity for binding sites in the zeroth to 57th
  percentile range (compared to $31\%$ faster for the 20th to 30th
  percentile range).}

To test whether the minimum for binding sites in euchromatin regions
was dependent on the spatial structure of the LAC distribution we
randomly shuffled the values of the LAC distribution among the voxels
of the 09 cell nucleus. This preserved the overall distribution of LAC
values, shown in Figure~\ref{fig:09-LAC}, while removing all spatial
correlations between the values in neighboring voxels.  As seen in
Figure~\ref{fig:randLocMean09}, for binding sites in euchromatin (the
20th to 30th percentile range), the occurrence of a minimum mean
binding time for non-zero volume exclusivity is lost (magenta curve).
We discuss this result further in the next section, where we show the
appearance of a minimum is to be expected if the potential is
\emph{slowly varying} relative to the length scale of the binding
site.

In Figure~\ref{fig:randLocCDFs} we show the survival distributions,
$\prob \brac{\tau > t}$, for the 09 cell nucleus when for each simulation
the binding site was randomly localized in regions of euchromatin
(20th to 30th percentile range). Note that for large values of the
volume exclusivity the survival probability is no longer
well-approximated by an exponential distribution, in contrast to the
case in which the binding site was fixed across all simulations (as in
Figure~\ref{fig:fixedLocCDF}). This non-exponential behavior arises
because the binding site location is now itself a random variable.
Figure~\ref{fig:randLocCDF09Zoom} uses an expanded $t$-axis from
Figure~\ref{fig:randLocCDF09} to illustrate how the survival
distribution shifts as the volume exclusivity is increased. We see
that initially the distribution shifts to the left, leading to faster
binding times, but that as $\phimax \to \infty$ the distribution
rapidly shifts rightward.

\begin{table} 
  \setlength{\tabcolsep}{10pt}
  \begin{tabular}{l|r|r|r|r|r|r} %
    Nucleus & 09 & 02 & 05-14-1 & 05-14-2 & 05-15 & SIM nucleus\\ \hline
    Percent Speedup & 31\% & 34\% & 24\% & 23\% & 28\% & 33\% \\
  \end{tabular}
  \caption{\small Percent difference between mean binding time with no 
    volume exclusion ($\phimax = 0$) and the smallest mean binding
    time as $\phimax$ is varied (from Figure~\ref{fig:randLocMeanMultcells}).
    The last column gives the percent speed up in the median binding time
    we found for a single mouse myoblast cell nucleus reconstructed from structured 
    illumination microscopy (SIM) of DAPI stained DNA~\cite{IsaacsonPNAS2011}.
    The corresponding speedup in median binding times for the X-ray tomography
    data we study here  was comparable
    to the values reported for the means above.}
  \label{tab:speedups}
\end{table}
The results of this section are consistent with those we found when
studying a single cell in~\cite{IsaacsonPNAS2011} using a chromatin
density field reconstructed from fluorescence imaging of DAPI stained
DNA. Our new simulations illustrate that across several cells of
different phenotypes these results persist. The use of X-ray
tomography provides a more direct measure of chromatin density since
each voxel's LAC is directly proportional to the density of organic
material within that voxel~\cite{McDermott:2009ib}. Thus, we would
expect the present results to capture more accurately the true spatial
distribution of chromatin within each nucleus. Across the five nuclei
of this paper, we see that for binding sites localized in euchromatin
volume exclusion can lead to $23\%$ to $34\%$ lower (\textit{i.e.},
faster) mean search times. The maximum observed speedups are
summarized in Table~\ref{tab:speedups}.

\section{Why Does the Mean Binding Time Exhibit a Minimum for Binding Sites in Euchromatin?} 
\label{S:mtTheory}

In this section we investigate why, for binding sites localized in
regions of euchromatin, the mean binding time has a minimum as
$\phimax$ is increased from zero to infinity.  Our analysis is based
on the assumption that near the binding site the volume exclusion
potential varies sufficiently slowly that it is well-approximated by a
constant. 

To estimate the solution to~\eqref{eq:rdme} we make several
assumptions:
\begin{enumerate}
\item There exists a collection of voxels, $I'$, about and including
  the binding site where the potential can be approximated as
  constant.
\item The portion of the nucleus given by the voxels in $I'$ is large
  in diameter relative to the length of one voxel, but small relative
  to the diameter of the entire nucleus.
\item The initial position of the protein is outside $I'$, that is
  $P_{\bi}(0) = 0$ for $\bi \in I'$.
\item The solution to~\eqref{eq:rdme} reaches a quasi-steady state in
  space within $I'$ faster than the binding time scale. We therefore
  assume that for $\bi \in I'$, $\D{P_{\bi}}{t} \approx 0$.
\item The time for the protein to find the binding site is
  sufficiently large that the solution to~\eqref{eq:rdme} outside $I'$
  is essentially in equilibrium \emph{before} the protein locates the
  binding site. In particular, all memory of the initial position is
  lost.
\end{enumerate}
Using these assumptions we construct two solutions to~\eqref{eq:rdme},
an inner solution valid near $\bib$, $\Pin_{\bi}(t)$, and an outer
solution valid outside $I'$, $\Pout_{\bi}(t)$. The last assumption
above implies that
\begin{equation} \label{eq:PoutSol}
  \Pout_{\bi}(t) = A(t) e^{-\phi_{\bi} / \kb T},  \quad \bi \in I - I',
\end{equation}
where $A(t)$ is a decreasing function of time, to be determined later.
Since, by assumption 1, $\phi_{\bi}$ is nearly constant on $I'$, we can 
use~\eqref{eq:PoutSol} to extend the definition of $\Pout_{\bi}$ so that
\begin{equation} \label{eq:PoutSolIp}
\Pout_{\bi}(t) = A(t) e^{-\phi_{\bib} / \kb T}, \quad \bi \in I'.
\end{equation}
The second, third, and fourth assumptions imply
\begin{equation}  \label{eq:PinNonConstPotEq}
  \sum_{\bj \in I} \alpha_{\bi \bj} \Pin_{\bj}(t) - \alpha_{\bj \bi} \Pin_{\bi}(t) = 0, 
  \quad \forall \bi \in I' - \{\bib\},
\end{equation}
with the boundary condition $\Pin_{\bib} = 0$. \comment{The assumption that the
potential is constant in $I'$ then
simplifies~\eqref{eq:PinNonConstPotEq} to
\begin{equation} \label{eq:PinEq}
 D \lap_{h} \Pin_{\bi}(t) =
 \frac{D}{h^2} \sum_{d=1}^3 \brac{ \Pin_{\bi + \ve_d}(t) + \Pin_{\bi - \ve_d}(t)  - 2 \Pin_{\bi}(t)}
 = 0, \quad \bi \in I' - \{\bib\},
\end{equation}
where $\lap_h$ denotes the discrete Laplacian and $\ve_d$ is a unit
vector along the $d$th coordinate axis of $\R^3$. This equation is
coupled to the boundary condition that $\Pin_{\bib}(t) = 0$.  Finally,
we make the standard inner solution assumption
(see~\cite{ColeKevorkianBook,WardCheviakov2011}) that $I'$ is
sufficiently large that the solution to this equation can be
approximated by the solution when $I' = \Z^3$.

If $A(t)$ were known, a unique solution to~\eqref{eq:PinEq} when
$I'=\Z^3$ could be obtained by specifying the matching condition that }
\begin{equation} \label{eq:matchCondit} 
  \lim_{\abs{\bi} \to \infty} \Pin_{\bi}(t) = \Pout_{\bib}(t).
\end{equation}
This condition assumes that $I'$ looks like a single voxel on length
scales of relevance to the outer solution.  We note that all
time-dependence in~\eqref{eq:PinEq} arises from this matching
condition.

To solve~\eqref{eq:PinEq} we consider a related problem where the absorbing 
boundary condition $\Pin_{\bib}(t) = 0$ is replaced with an explicit
sink. Let $C_{\bi}(t)$ satisfy 
\begin{equation} \label{eq:cEq}
  D \lap_h C_{\bi}(t) = K(t) \delta_{\bi \vec{0}}, \quad \bi \in \Z^3,
\end{equation}
with $\lim_{\abs{\bi} \to \infty} C_{\bi}(t) = 0$. Here $K(t)$ denotes
the probability per time the protein is removed from the origin at
time $t$.  The choice $\Pin_{\bi}(t) = C_{\bi - \bib}(t) - \CO(t)$
satisfies~\eqref{eq:PinEq} with the boundary condition $\Pin_{\bib}(t)
= 0$. Note that $C_{\bi}(t) \leq 0$, with its most negative value at
$\bi = \vec{0}$.  Therefore $\Pin_{\bi}(t) \geq 0$, as required.  The
matching condition~\eqref{eq:matchCondit} then gives
\begin{equation*}
\lim_{\abs{\bi} \to \infty} \Pin_{\bi}(t) = -\CO(t) = A(t) e^{-\phi_{\bib} / \kb T},
\end{equation*}
so that solving for $A(t)$, and substituting the result
into~\eqref{eq:PoutSol} and~\eqref{eq:PoutSolIp} we find
\begin{equation} \label{eq:pOutSolut}
  \Pout_{\bi}(t) = -\CO(t) e^{\paren{\phi_{\bib} - \phi_{\bi}}/ \kb T}.
\end{equation}

$\CO(t)$ can be determined using Fourier transforms. Let $B =
\brac{-\pi,\pi}^{3}$ denote the cube centered at the origin with edges
of length $2 \pi$. Then for $\vxi = (\xi_1,\xi_2,\xi_3)$ labeling a
point in $B$,
\begin{equation*}
  C_{\bi}(t) = \frac{1}{(2 \pi)^{3}} \int_{B} \hat{C}(\vxi,t) \, e^{i (\bi \cdot \vxi)} d \vxi,
\end{equation*}
where $\hat{C}(\vxi,t)$ denotes the Fourier transform of $C_{\bi}(t)$
and $i = \sqrt{-1}$. The solution to~\eqref{eq:cEq} in Fourier space is
easily found to be 
\begin{equation*}
  \hat{C}(\vxi,t) = -\frac{K(t) h^2}{4 D} \frac{1}{\sum_{d=1}^{3} \sin^2 \paren{\frac{\xi_d}{2}}},
\end{equation*}
so that
\begin{align}
  \CO(t) &= - \frac{K(t) h^2}{32 \pi^3 D} \int_{B} \frac{1}{\sum_{d=1}^{3} \sin^2 \paren{\frac{\xi_d}{2}}} \, d \vxi , \notag \\
  &= - \frac{2 K(t) h^2}{\pi^3 D} \int_{\tilde{B}} \frac{1}{\sum_{d=1}^{3} \sin^2 \paren{\xi_d}} \, d \vxi, \label{eq:C0Val} 
\end{align}
where $\tilde{B} = \brac{0,\frac{\pi}{2}}^3$. We note that the
singularity of the integrand in~\eqref{eq:C0Val} is like
$\abs{\vxi}^{-2}$, and hence integrable.

To find the distribution of binding times, and in particular the mean
binding time, we make the approximation that the survival probability
at any given time $t$ can be found from the outer solution alone,
\textit{i.e.}, we assume that
\begin{equation*}
\prob \brac{ \tau >t } = \sum_{\bi \in I} \Pout_{\bi}(t) = - \CO(t) e^{-\phi_{i_b}/\kb T} Z,
\end{equation*}
where $Z$ is the partition function
\begin{equation*}
 Z = \sum_{\bi \in I} e^{-\phi_{\bi} / \kb T}. 
\end{equation*}
If we substitute $\CO(t)$ as given by~\eqref{eq:C0Val} into the above
equation, and if we also recall that $K(t)$ is the probability per
unit time of binding, and therefore that
\begin{equation*}
K(t) = - \D{}{t} \prob \brac{ \tau >t }                            
\end{equation*}
we get
\begin{equation} \label{eq:survProbTheory}
\prob \brac{\tau > t} = - \avg{\tau} \D{}{t} \prob \brac{\tau > t}  
\end{equation}
where
\begin{equation}  \label{eq:mfptLat}
\avg{\tau} = \frac{2}{\pi^3 D h} \paren{\sum_{\bi\in I} e^{\paren{\phi_{\bib} - \phi_{\bi}}/\kb T} h^3}  \int_{\tilde{B}} \frac{1}{\sum_{d=1}^{3} \sin^2 \paren{\xi_d}} \, d \vxi.
\end{equation}
Equation~\eqref{eq:survProbTheory} shows that the distribution of
binding times is exponential, and~\eqref{eq:mfptLat} gives an explicit
formula for the mean binding time, $\avg{\tau}$.  Note that the
theoretical estimate~\eqref{eq:mfptLat} involves no unknown
parameters. No parameter fitting is necessary to compare the
predicted mean binding time~\eqref{eq:mfptLat} to the simulation
results of the previous section.

As the voxel size approaches zero, $\avg{\tau} \to \infty$ like
$h^{-1}$.  This is consistent with the well-known fact that it takes
an infinite amount of time for a point particle to find a point target
by diffusion in a three-dimensional space.  (Recall that our target is
one voxel in all of the computations reported here.)  As we approach
the limit $h \to 0$, the last of our simplifying assumptions becomes
more and more valid, since there is more and more time for
$\Pout_{\bi}$ to equilibrate before binding occurs.  Related to this,
the assumption that the initial position of the protein does not
matter also becomes better and better as $h \to 0$.

\begin{figure}
  \centering
  \begin{minipage}{.49\columnwidth}
    \subfloat[]{
      \label{fig:mtTheoryFixedTarg09} 
      \scalebox{.42}{\includegraphics{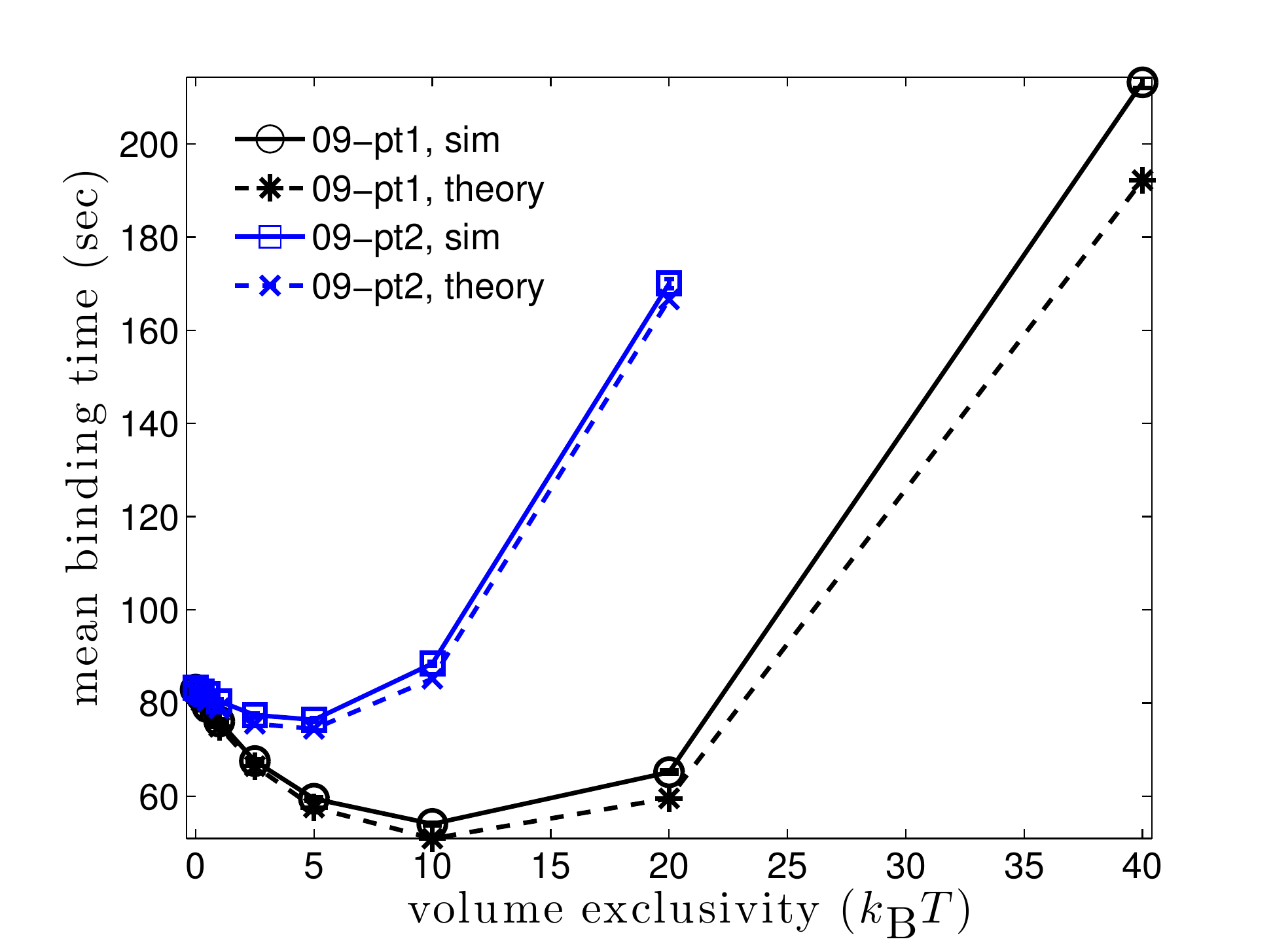}}}  \\
    \vspace{-10pt}
    \subfloat[]{
      \label{fig:mtTheoryRandTargs02} 
      \scalebox{.42}{\includegraphics{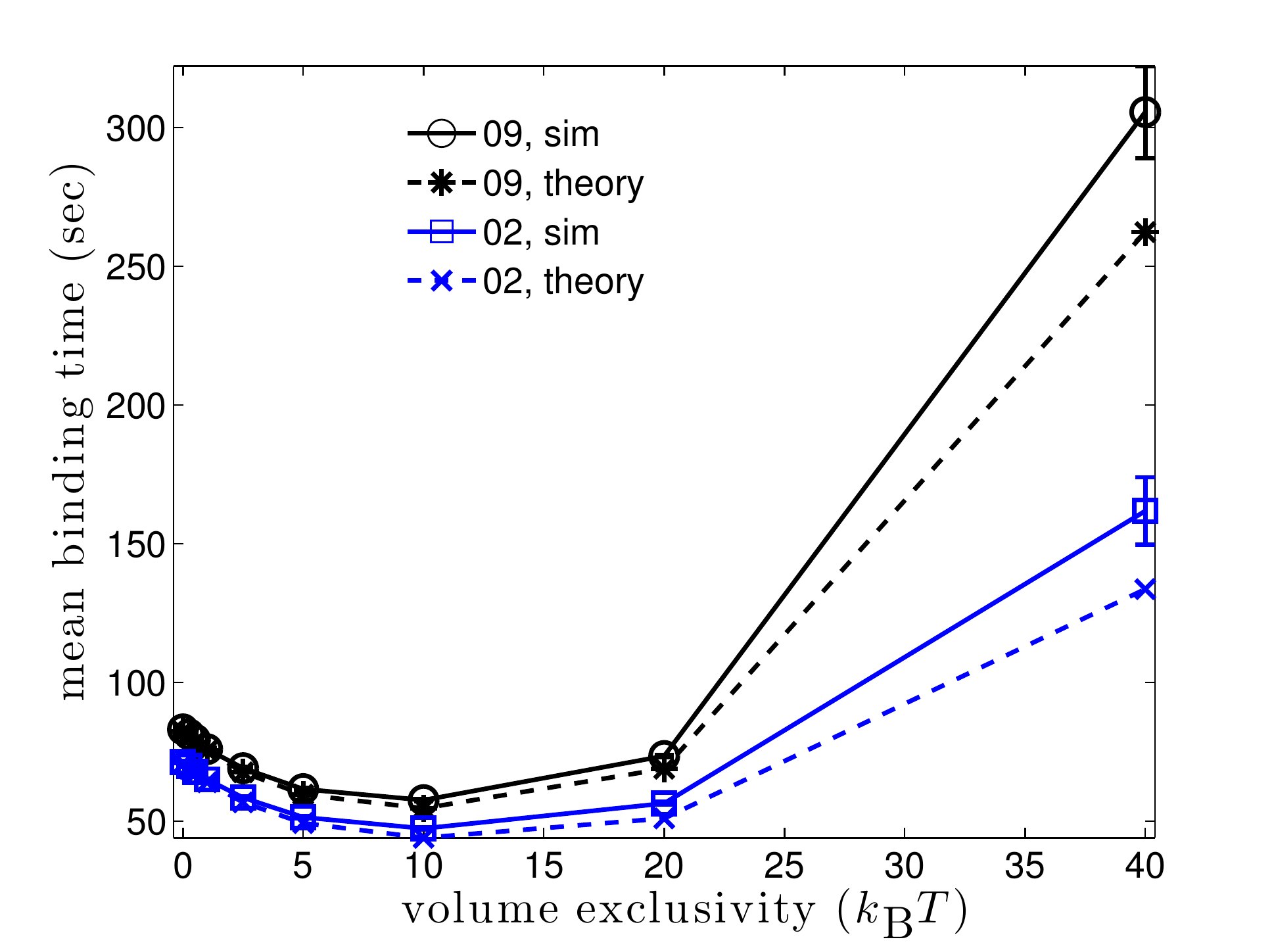}}}
  \end{minipage}
  \hspace{-0.04\linewidth}
  \begin{minipage}{.49\columnwidth}    
  \subfloat[]{
    \label{fig:meanTheoryRandTargs} 
    \scalebox{.42}{\includegraphics{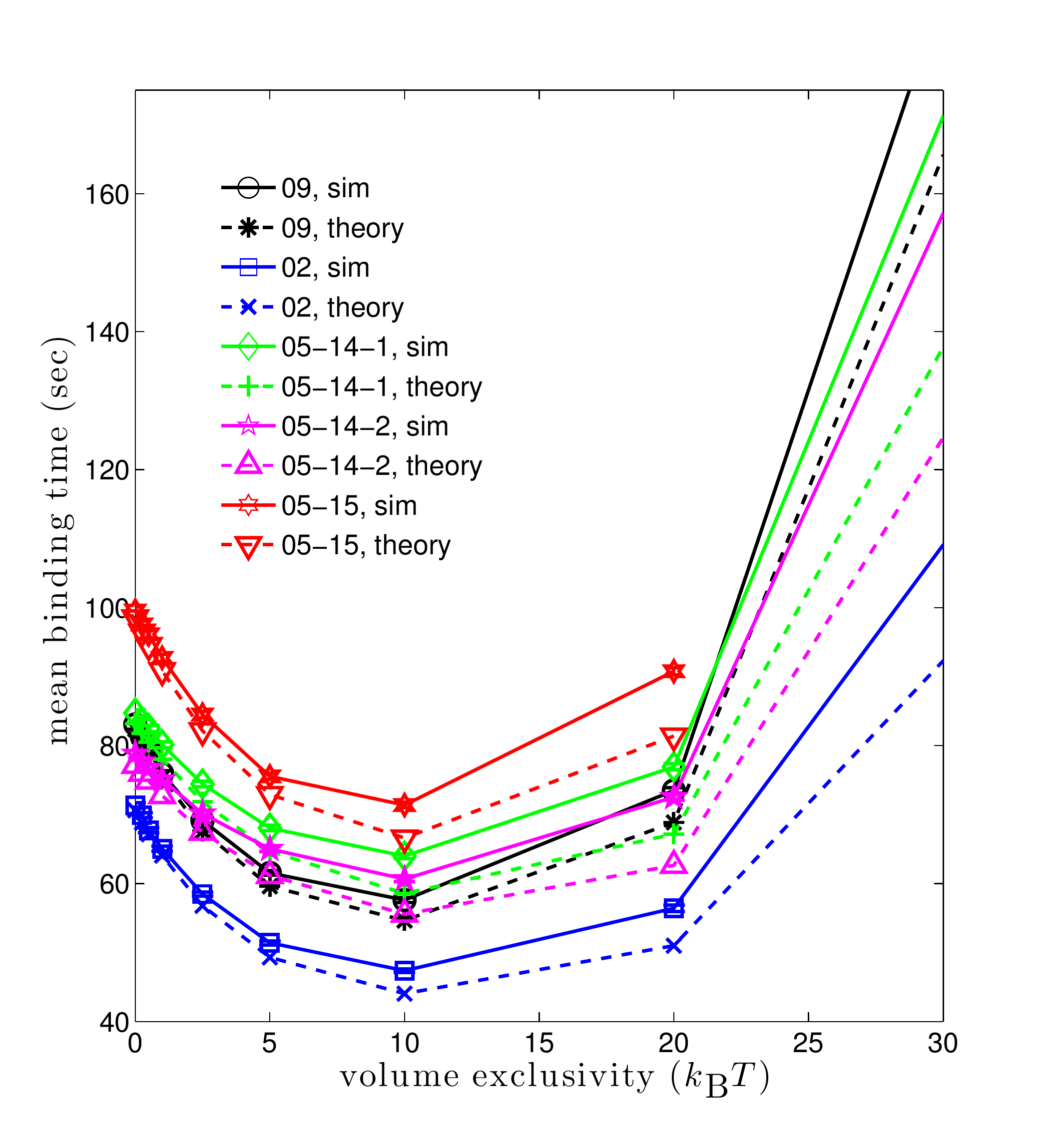}}}
  \end{minipage}
  \caption{\small Theoretical mean binding times~\eqref{eq:mfptLat}
    and~\eqref{eq:meanRandTargs} versus those from simulations. Note
    that both theoretical mean binding times involve no fitting to the
    simulated times, and are completely determined by known
    parameters.  In all figures solid lines give the mean binding
    times from simulations, while dashed lines give the corresponding
    theoretical prediction.  (a) 09 cell for the two fixed targets
    used in Figure~\ref{fig:fixedLocMean}.  Theoretical mean binding
    times are from~\eqref{eq:mfptLat}. (b) 09 and 02 cells with
    binding sites randomly localized in euchromatin (20th to 30th
    percentile of LAC distribution). Simulated means are those from
    Figure~\ref{fig:randLocMeanMultcells}, while theoretical are
    from~\eqref{eq:meanRandTargs}. (c) All five nuclei when binding
    sites are randomly localized in regions of euchromatin (20th to
    30th percentile of the LAC distribution).  Simulated means are the
    same is in Figure~\ref{fig:randLocMeanMultcells}, while
    theoretical are from~\eqref{eq:meanRandTargs}.}
  \label{fig:meanTheory}
\end{figure}
In Figure~\ref{fig:mtTheoryFixedTarg09} we compare the mean time to
find a target binding site predicted by~\eqref{eq:mfptLat} to the
empirical mean times found from the simulations in
Figure~\ref{fig:fixedLocMean}.  The solid curves in
Figure~\ref{fig:mtTheoryFixedTarg09} show the simulated mean binding
times for the two target locations in the 09 cell that were used in
Figure~\ref{fig:fixedLocMean}. The dashed curves show the estimated
mean binding times using~\eqref{eq:mfptLat}.  The theory gives very
good agreement in the predicted mean binding time, particularly for
smaller values of the volume exclusivity, $\phimax$.

To see why the mean binding time has a minimum for binding sites in
euchromatin regions we look at the behavior of~$\avg{\tau}$ as $\phimax$
is increased from zero. Let
\begin{equation*}
  \hat{\tau} = \frac{2 h^2}{\pi^3 D} \int_{\tilde{B}} \frac{1}{\sum_{d=1}^{3} \sin^2 \paren{\xi_d}} \, d \vxi.
\end{equation*}
Using~\eqref{eq:potentDef} we have that 
\begin{equation} \label{eq:meanTimeVolExFunc}
  \avg{\tau} = \hat{\tau} \sum_{\bi\in I} e^{\phimax \paren{\ell_{\bib} - \ell_{\bi}} / \kb T},
\end{equation}
where $\ell_{\bi}$ denotes the normalized LAC in voxel $\bi$. We now
consider the mean binding time as a function of $\phimax$,
$\avg{\tau}(\phimax)$.  Taking the derivative in $\phimax$,
\begin{equation*}
  \D{}{\phimax} \avg{\tau}(\phimax) = \frac{\hat{\tau}}{\kb T} \sum_{\bi\in I} \paren{\ell_{\bib} - \ell_{\bi}} e^{\phimax \paren{\ell_{\bib} - \ell_{\bi}} / \kb T}.
\end{equation*}
At $\phimax = 0$, $\avg{\tau}(\phimax)$ will be decreasing when
$\ell_{\bib}$ is smaller than the mean of the nuclear LAC
distribution, and increasing when $\ell_{\bib}$ is above the mean.
Examining the second derivative we see
\begin{equation*}
  \DD{}{\phimax} \avg{\tau}(\phimax) = \frac{\hat{\tau}}{(\kb T)^2} \sum_{\bi\in I} \paren{\ell_{\bib} - \ell_{\bi}}^2 e^{\phimax \paren{\ell_{\bib} - \ell_{\bi}}/\kb T} > 0
\end{equation*}
if the LAC values are non-constant. We therefore find that the first
derivative is a strict monotone increasing function. 

Note that 
\begin{equation} \label{eq:expLim}
  \lim_{\phimax \to \infty} e^{\phimax \paren{\ell_{\bib} - \ell_{\bi}}/ \kb T} = 
  \begin{cases}
    0,  &\ell_{\bib} < \ell_{\bi},\\
    1,  &\ell_{\bib} = \ell_{\bi},\\
    \infty,  &\ell_{\bib} > \ell_{\bi}.
  \end{cases}
\end{equation}
If $\ell_{\bib} \neq \min_{\bi \in I} \ell_{\bi}$, there is at least
one term in the sum~\eqref{eq:mfptLat} for which $\ell_{\bib} -
\ell_{\bi} > 0$. As such,
\begin{equation*}
  \lim_{\phimax \to \infty} \avg{\tau}(\phimax) = \infty.
\end{equation*}

The preceding results allow for a possible explanation of the
dependence on $\phimax$ of the mean binding times observed in our
simulations.  For binding sites in regions below the mean of the LAC
distribution, the mean binding time will always decrease to a minimum
as $\phimax$ is increased from zero.  Beyond this minimum the mean
binding time will increase to infinity as $\phimax \to \infty$.  In
contrast, for binding sites above the mean of the LAC distribution the
mean binding time will simply increase as $\phimax$ is increased.
By~\eqref{eq:pOutSolut} we see that $\Pout_{\bi}(t) \propto
\exp\brac{\phimax ( \ell_{\bib} - \ell_{\bi})/\kb T}$. Combined
with~\eqref{eq:expLim} this suggests that as $\phimax$ is increased
from zero the effective volume the protein must explore is decreased,
while the potential barriers provided by regions of high LACs the
protein must cross to move between two regions of low LAC become
higher and higher. For binding sites in euchromatin there is a balance
between the two effects at which the mean binding time is minimized.
In contrast, for binding sites in heterochromatin the latter effect
wins out, and the mean binding time simply increases as $\phimax$
increases.

We now consider how the mean binding time behaves when the binding
site is itself randomly localized within subregions of the nucleus.
Let $\avg{\Teuc}$ denote the mean binding time when the binding site
position is chosen from a uniform distribution among voxels within a
region of euchromatin, $I_{\textrm{euc}} \subset I$. We will
subsequently take $I_{\textrm{euc}}$ to be the collection of all
voxels of a nucleus containing euchromatin within the 20th to 30th
percentiles of the nuclear LAC distribution. If $V_{\textrm{euc}}$
denotes the total volume of the voxels within $I_{\textrm{euc}}$,
\begin{align}
  \avg{\Teuc} &= \frac{1}{V_{\textrm{euc}}} \sum_{\bib \in I_{\textrm{euc}}} \avg{\tau} h^3, \notag \\
  &=  \frac{2}{\pi^3 D h V_{\textrm{euc}} } \paren{\sum_{\bi\in I} \sum_{\bib \in I_{\textrm{euc}}} e^{\paren{\phi_{\bib} - \phi_{\bi}}/\kb T} h^6}  \int_{\tilde{B}} \frac{1}{\sum_{d=1}^{3} \sin^2 \paren{\xi_d}} \, d \vxi. 
  \label{eq:meanRandTargs}
\end{align}
As in~\eqref{eq:mfptLat}, we see that~\eqref{eq:meanRandTargs}
completely specifies $\avg{\Teuc}$ in terms of known parameters.

Figure~\ref{fig:mtTheoryRandTargs02} compares~\eqref{eq:meanRandTargs}
to the mean times from simulations in the 09 and 02 cell nuclei, while
Figure~\ref{fig:meanTheoryRandTargs} compares~\eqref{eq:meanRandTargs}
to simulations for all five cell nuclei (see
Figure~\ref{fig:randLocMeanMultcells}).  In every nucleus we see that
for $\phimax \leq 10 \, \kb T$ the theory gives good estimates of the
mean binding time within the nuclei. As $\phimax \to \infty$ the
theoretical formula~\eqref{eq:meanRandTargs} appears to underestimate
the mean binding time found by the SSA simulations. This breakdown
could arise for several reasons. Foremost, as $\phimax$ is increased
the difference between potential values in neighboring voxels
increases. The assumption that the potential is approximately constant
near the target binding site can therefore break down. 

It should be noted that both~\eqref{eq:mfptLat}
and~\eqref{eq:meanRandTargs} depend only on the value of the target
binding site LAC relative to the mean LAC, and not on the detailed
spatial structure of the LAC distribution.  Based on this observation,
one might wonder why we do not observe the same dependence of the mean
binding time on $\phimax$ when the values of the LAC distribution are
randomly shuffled among voxels (see Figure~\ref{fig:randLocMean09}).
The answer is that the conditions under which~\eqref{eq:mfptLat}
and~\eqref{eq:meanRandTargs} were derived no longer hold. After randomly
shuffling the LAC distribution the potential can no longer be
approximated as constant near a binding site.


\section{Conclusions}

We have applied the model we developed in~\cite{IsaacsonPNAS2011} to
study how the time required for proteins to find a specific binding
site varies as a function of volume exclusion by dense regions of
chromatin and binding site localization. Linear absorption
coefficients from soft X-ray tomography reconstructions of five mouse
olfactory sensory neurons were used to determine the spatial variation
of chromatin density within nuclei. The distribution of LACs within
each of the nuclei was observed to be bimodal, demonstrating the
spatial separation of nuclear space into regions of euchromatin, where
most active genes are localized, and denser heterochromatin, where
silenced genes are typically located.

Numerical simulations of our model suggest that for binding sites
localized in regions of euchromatin there exists a non-zero volume
exclusivity at which the mean binding time is minimized.  As the
volume exclusivity was increased beyond this minimum the mean binding
time simply increased to infinity. Across the five nuclei used in this
study, the minimal mean binding time was found to be $23$ to $34$
percent faster than that observed in simulations where the volume
exclusivity was zero (\textit{i.e.} the protein simply diffused,
experiencing no volume exclusion).  Randomly shuffling the LAC values
among the voxels of the nucleus led to a loss of this minimum,
suggesting that the spatial distribution of the chromatin plays a roll
in the existence of a minimal mean binding time for non-zero volume
exclusivity.  For binding sites localized in heterochromatin the
mean binding time simply increased as the volume exclusivity was
increased from zero.

The observed behavior for binding sites localized in either
euchromatin or heterochromatin can be explained by the analytical
formulas~\eqref{eq:mfptLat} and~\eqref{eq:meanRandTargs}.  These
approximations to the mean binding time were derived under several
assumptions, including that the LAC values in voxels near a binding
site are approximately constant and that the time to find the binding
site is sufficiently large that the distribution of the protein's
position is proportional to the equilibrium Gibbs-Boltzmann
distribution~\eqref{eq:rdmeSS}. Assuming these conditions hold,
both~\eqref{eq:mfptLat} and~\eqref{eq:meanRandTargs} suggest that the
observed dependence of the mean binding time on the volume exclusivity
is determined by whether the binding site LAC is below or above the
mean LAC within the nucleus. For binding sites with LACs below the
mean, such as those localized in euchromatin, the theory predicts the
appearance of a minimum mean binding time for non-zero values of the
volume exclusivity. It appears that increasing the volume exclusivity
from zero helps speed up the search process by decreasing the
effective volume that must be searched to find the binding site.
Beyond the value that minimizes the mean binding time, further
increasing the volume exclusivity leads to increased binding times as
the protein becomes trapped in regions surrounded by steep potential
barriers.  For binding sites localized in regions above the mean, such
as heterochromatin, the theory agrees with the observed dependence in
our simulations, predicting the mean binding time will simply increase
as the volume exclusivity increases from zero.

It should be noted that our theory does not give an explanation for
the dependence of the mean binding time on the volume exclusivity when
the LAC values are shuffled. In this case the assumption that the LAC
values near the binding site are approximately constant is violated,
so that~\eqref{eq:mfptLat} and~\eqref{eq:meanRandTargs} no longer
hold. The simulations with the shuffled LAC values, combined with our
analytical theory, suggest that a key aspect of the macroscopic
spatial distribution of chromatin that could lead to a decreased mean
binding time caused by volume exclusion is a slow variation in chromatin
density in the neighborhood of binding sites localized in euchromatin.

\begin{acknowledgments}
  SAI, DMM, and CSP were supported by the Systems Biology Center New
  York (National Institutes of Health Grant P50GM071558). SAI was also
  supported by National Science Foundation grant DMS-0920886. MLG and
  CAL were supported by the Department of Energy Office of Biological
  and Environmental Research Grant DE-AC02-05CH11231, the NIH National
  Center for Research Resources (5P41 RR019664-08) and the National
  Institute of General Medical Sciences (8P41 GM103445-08) from the
  National Institutes of Health.
\end{acknowledgments}

\appendix

\comment{
  \section{Soft X-ray Tomography Measurement Error} \label{ap:measErr}
  The X-ray microscope employs monochromatic X-rays and therefore the
  values obtained from computed tomography measurements are equal to
  the LAC values calculated from the atomic composition of the
  specimen. The SXT technique avoids the beam hardening effects
  commonly found in polychromatic tomographic imaging
  (see~\cite{Tsuchiyama2005tj}). The measurement error for each pixel
  of a single projection image is of order 3\%, determined by photon
  shot noise.  The LAC value of each 32nm voxel is obtained from
  tomographic reconstruction of many such projections and is typically
  less than 1\%.  LAC measurement errors are insignificant compared to
  the observed cell-to-cell variation.  }

\bibliographystyle{plain}


\end{document}